\documentclass[
    aps,
    prx,
    letterpaper,
    nobalancelastpage,
    twocolumn,
    superscriptaddress,
    nofootinbib,
    longbibliography
]{revtex4-2}

\usepackage{graphicx}
\usepackage{amsmath}
\usepackage{amssymb}
\usepackage[english]{babel}
\usepackage{color}
\usepackage[version=4]{mhchem}
\usepackage[hidelinks]{hyperref}
\usepackage{dsfont}
\usepackage{multirow}
\usepackage{booktabs}
\usepackage{array}
\usepackage{subcaption}

\usepackage{soul}
\newcommand{\avg}[1]{\left< #1 \right>}

\newcommand{\ket}[1]{{| {#1} \rangle}}
\newcommand{\bra}[1]{{\langle {#1} |}}
\newcommand{\braket}[2]{\langle {#1} | {#2} \rangle}

\newcommand{\UIUC}{
    Department of Physics,
    The University of Illinois at Urbana-Champaign,
    Urbana, IL 61801, USA
}
\newcommand{\Harvard}{
    Department of Physics,
    Harvard University,
    Cambridge, MA 02138, USA
}
\newcommand{\Stockholm}{
    Department of Physics,
    Stockholm University,
    SE-106 91 Stockholm, Sweden
}
\newcommand{\Stevens}{
    Department of Physics,
    Stevens Institute of Technology,
    Hoboken, NJ 07030, USA
}

\begin{document}

\title{Probing curved spacetime with a distributed atomic processor clock}
%\author{People}

\author{Jacob P. Covey}\email{jcovey@illinois.edu}
\affiliation{\UIUC}
\author{Igor Pikovski}\email{pikovski@stevens.edu}
\affiliation{\Stockholm}
\affiliation{\Stevens}
\author{Johannes Borregaard}\email{borregaard@fas.harvard.edu}
\affiliation{\Harvard}

\begin{abstract}
Quantum dynamics on curved spacetime has never been directly probed beyond the Newtonian limit. Although we can describe such dynamics theoretically, experiments would provide empirical evidence that quantum theory holds even in this extreme limit. The practical challenge is the minute spacetime curvature difference over the length scale of the typical extent of quantum effects. Here we propose a quantum network of alkaline earth(-like) atomic processors for constructing a distributed quantum state that is sensitive to the differential proper time between its constituent atomic processor nodes,  implementing a quantum observable that is affected by post-Newtonian curved spacetime. Conceptually, we delocalize one clock between three locations by encoding the presence or absence of a clock into the state of the local atoms. By separating three atomic nodes over $\sim$km-scale elevation differences and distributing one clock between them via a W-state, we demonstrate that the curvature of spacetime is manifest in the interference of the three different proper times that give rise to three distinct beat notes in our non-local observable. We further demonstrate that $N$-atom entanglement within each node enhances the interrogation bandwidth by a factor of $N$. We discuss how our system can probe new facets of fundamental physics, such as the linearity, unitarity and probabilistic nature of quantum theory on curved spacetime. Our protocol combines several recent advances with neutral atom and trapped ions to realize a novel quantum probe of gravity uniquely enabled by quantum networks. 
\end{abstract}
\maketitle

\section{Introduction}

The interface between quantum theory and gravity remains elusive. There are two main manifestations of this interface: the quantization of the gravitational field itself \cite{oriti2009approaches}, and the dynamics of quantum systems on a classical curved spacetime \cite{fulling1989aspects}. Despite recent ideas on detecting quantum features of gravity with novel quantum systems \cite{carney2019tabletop, bose2023massive, marletto2024quantum, lami2024testing, tobar2024detecting}, their experimental implementation remains a significant challenge. In contrast, detecting the influence of curved spacetime on quantum systems only requires quantum mechanical probes. The effect of Earth's gravity in the dynamics of quantum systems can be readily observed. But so far, this has only been achieved in the Newtonian limit, such as in matter-wave interferometry \cite{colella1975observation,overstreet2022observation} where Newtonian gravity induces a coherent phase-shift, or in neutron bouncing experiments \cite{nesvizhevsky2002quantum,jenke2009q} where the Newtonian potential acts as a potential well for neutrons. Going beyond the Newtonian limit, and observing genuine quantum phenomena affected by curved spacetime, remains a challenge. While post-Newtonian corrections to the potential can result in modifications of the phase-shift \cite{dimopoulos2008general}, there are also effects unique to gravity, such as gravitational time dilation, which result in different signatures in the quantum domain.

\begin{figure}[t!]
    \centering
    \includegraphics[width=0.48\textwidth]{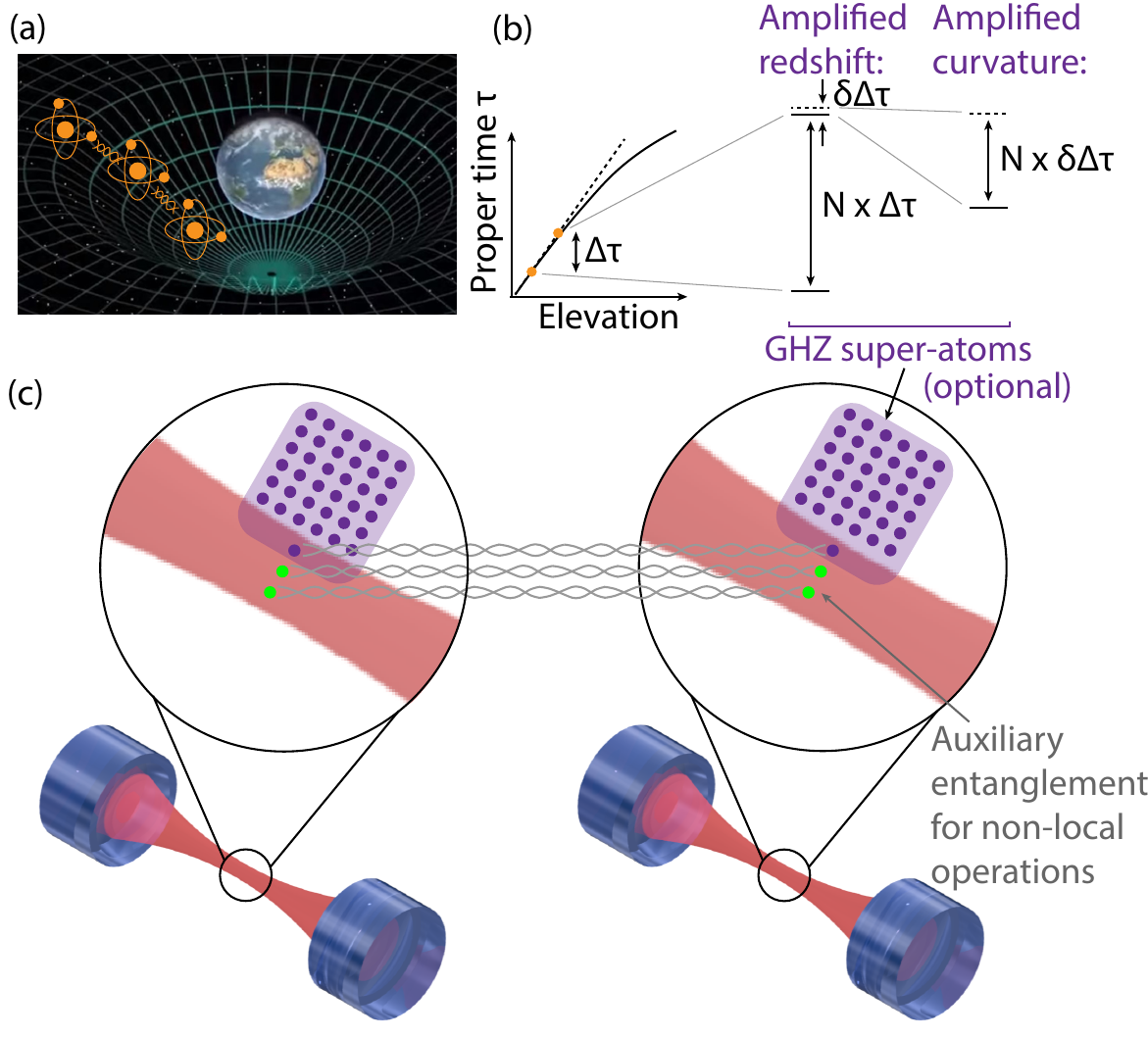}
    \caption{
        \textbf{Probing curved spacetime with entangled clocks.} (a) Three atomic clock systems in locations that experience different local gravity share a W-state. (b) The GHZ super-atoms can boost the bandwidth by amplifying the general-relativistic proper time difference $\Delta\tau$ and nonlinearity $\delta\Delta\tau$ between different elevations, but are otherwise not necessary. (c) The experimental approach. Atomic processor clocks in optical cavities that include: the GHZ super-atom (purple dots) and auxiliary single-atom Bell pairs for non-local operations (green dots).
        \label{Figure1}
    }
\end{figure}

One such effect is the loss and revival of coherence due to gravitational time dilation, for which both quantum theory and general relativity have to be taken into account \cite{Zych2011}. This can be achieved through quantum interference of clocks that experience different proper times in superposition \cite{Zych2011,sinha2011atom}. One prepares internal states in superposition, such as a qubit, and also separates the system into a superposition of two or more distant locations. The time-dilation induced by Earth's gravitational field will cause the delocalized clock qubit to evolve differently in the two positions. This results in a periodic loss and revival of the visibility of interferometric ``self-interference" measurements as a signature of the interplay of quantum dynamics and general relativistic proper time evolution \cite{Zych2011, Pikovski2015,zych2016general,orlando2017does,zhou2018quantum,loriani2019interference,roura2020gravitational}. While the predictions have been simulated with BECs in a magnetic field \cite{margalit2015self}, the experiment in Earth's gravitational field remains challenging due to the required separations and coherence times: to see full loss and revival, one needs a spatial superposition on the order of $10$~m and coherence of $1$~s, in addition to the preparation of both internal and external superpositions in the same setup \cite{Zych2011}. 

An alternative route has recently been proposed that relies on entanglement, rather than matter-wave interferometry: one can encode the clocks in photons with different frequencies and use quantum memories \cite{barzel2024entanglement}, or use entangled networks of clocks to accumulate and then interfere different proper time evolutions~\cite{borregaard2024testing}. The use of quantum networks significantly increases the possible separations of the probe systems, and thus also opens the possibility to test the coherent dynamics of quantum systems on curved spacetime. However, the test proposed in Ref.~\cite{borregaard2024testing} does not explicitly probe the effect of the curvature of spacetime since it only involves the interference of two points in space. Measurements of spacetime curvature (beyond Newtonian gravity) on quantum systems will require distributing the quantum clocks over at least three locations. Such measurements are of particular importance to test quantum theory in this unexplored domain, as curved spacetime may result in modifications to the expected dynamics. However, since the nonlinearity is a higher-order effect in the gravitational potential, more stringent requirements on clock separations and coherence times are necessary to explore this regime.

Here, we propose to use a distributed atomic quantum processor with built-in optical atomic clock functionality to overcome these challenges (see Fig.~\ref{Figure1}). Specifically, we consider three atomic array quantum processors that are capable of distributing Bell pairs between them via photonic interconnects. We propose a realistic implementation of this system using arrays of ytterbium-171 atoms in optical cavities as network nodes. We demonstrate how this platform can be used to directly measure the curvature of spacetime via an interference fringe of a three-node entangled state with a $\sim$km-scale elevation difference between each node. In particular, we discuss how this enables first tests of the linearity and unitarity of quantum theory, and tests of the validity of Born's rule in the presence of relativistic spacetime curvature. While arguably challenging, the proposed setup is within reach of state-of-the-art hardware. We also discuss the prospect for employing non-classical states such as multi-atom Greenberger-Horne-Zeilinger (GHZ) states to effectively amplify the curvature of spacetime. 

\section{The atomic system and overview of the protocol}

\begin{figure}[t!]
    \centering
    \includegraphics[width=0.48\textwidth]{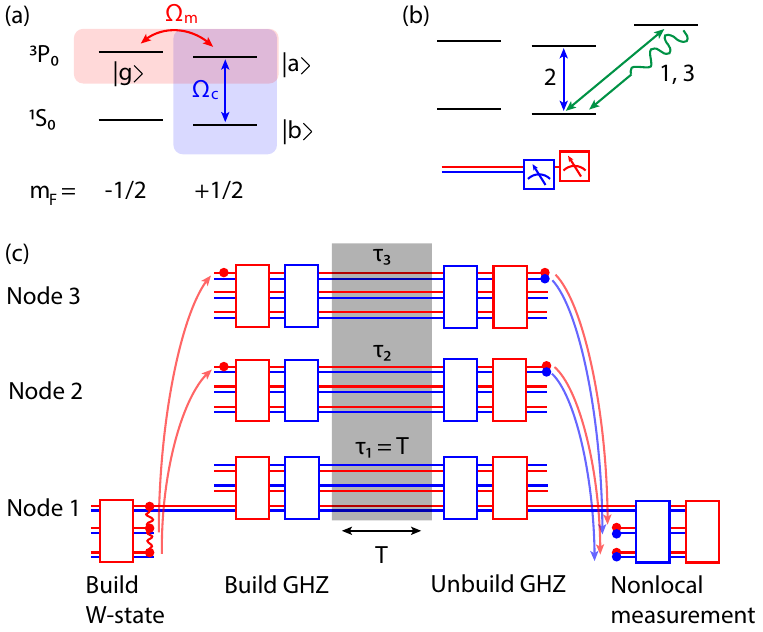}
    \caption{
        \textbf{Atomic system and protocol overview.} (a) The relevant level structure of $^{171}$Yb. The \{ga\}-sector (red) is defined by the nuclear spin qubit in the metastable $^3$P$_0$ state. The photon-mediated remote entanglement generation operates in this sector. The clock transition (blue) can be driven both to create coherence on the optical qubit \{ab\} ($\pi/2$-pulse) and also to swap the \{ga\}-sector for the \{gb\}-sector ($\pi$-pulse). (b) The three-state qutrit Hilbert space can be measured in two rounds of fluorescence detection in which one or both of the $^1$S$_0$ ground nuclear spin states is illuminated with probe light. The $^3$P$_0$ states remain dark and unaffected, and a clock $\pi$-pulse is used between the two rounds. (c) A schematic overview of the circuit. The W-state creation, manipulation, and detection is described in Fig.~\ref{Figure3}, while the GHZ state building and unbuilding is described in Fig.~\ref{Figure5}. 
        \label{Figure2}
    }
\end{figure}

Instead of delocalizing an atom to be in a superposition of three distant locations, we use three separated atomic systems and construct a collective quantum state in which we encode one ``clock" (``c") and two ``not clocks" (``nc")~\cite{borregaard2024testing}. We encode the ``nc" state and the initial part of the ``c'' state in the nuclear spin-1/2 degree of freedom of the metastable $^3$P$_0$ state of ytterbium-171 ($^{171}$Yb) such that the ``nc" state $|g\rangle$ is the $m_F=-1/2$ state while the initial ``c" state $|a\rangle$ is the $m_F=1/2$ state [see Fig.~\ref{Figure2}(a)]. We assume a heralded remote entanglement protocol that establishes the Bell state
\begin{equation}
|\psi\rangle=\frac{1}{\sqrt{2}}(|g,~a\rangle+|a,~g\rangle).
\end{equation}
Such protocols have been proposed in Refs.~\cite{Covey2019b,Huie2021,LiThompson2024,Sunami2024}, based on recent demonstrations of atom-photon entanglement and networked atom-atom entanglement of neutral alkali species~\cite{Hoffmann2012,Daiss2021,Covey2023,Hartung2024,Li2025}, trapped ions~\cite{Olmschenk2009,Stephenson2020,Nichol2022,Saha2024}, and solid-state spins~\cite{Bernien2013,Humphreys2018,Kindem2020,Bhaskar2020,Uysal2024}.

We refer to this qubit as ``metastable" (``m"), for which $|g\rangle$ and $|a\rangle$. The qubit energy is $E_m/h\approx100$ kHz at a modest field of $\approx100$ G ($\approx1$ kHz/G). We denote operations on this qubit by pulses with Rabi frequency $\Omega_m$ [see Fig.~\ref{Figure2}(a)] and color them red. Arbitrary fast and high-fidelity qubit manipulations have been demonstrated~\cite{Lis2023,Ma2023,Muniz2024}, and a coherence time of $\approx7$ seconds has been observed~\cite{Lis2023}. Additionally, high-fidelity, Rydberg-mediated two-qubit controlled-phase (CZ) gates have been demonstrated~\cite{Ma2023,Peper2024,Muniz2024}. We assume $|a\rangle$ is the only state coupled to a Rydberg state. Other two-qubit gates such as controlled-NOT (CNOT) gates can be constructed by sandwiching the CZ gate with single-qubits operation on the target qubit, such as with Hadamards in the case of the CNOT gate.

The optical atomic clock qubit is defined by $|a\rangle$, $m_F=1/2$ in the metastable $^3$P$_0$ state, and $|b\rangle$, $m_F=1/2$ in the ground $^1$S$_0$ state [see Fig.~\ref{Figure2}(a)]. The qubit energy is $E_c/h\approx520$ THz. We denote operations on this qubit by pulses with Rabi frequency $\Omega_c$ and color them blue. Arbitrary fast and high-fidelity qubit manipulations have been demonstrated on optical qubits~\cite{Lis2023,Finkelstein2024}, and a coherence time of $\approx30$ seconds has been observed for state-of-the-art optical clocks~\cite{Young2020,Bothwell2022}. Additionally, high-fidelity, Rydberg-mediated two-qubit controlled-phase (CZ) gates have been demonstrated on optical qubits~\cite{schine2022,Cao2024}. Again, we assume $|a\rangle$ is the only state coupled to a Rydberg state. Only a modest Zeeman splitting is required to completely decouple the $^1$S$_0$ $m_F=-1/2$ state from the other three, given the large differential g-factor between the metastable and ground nuclear spins~\cite{Lemke2012,Lis2023}. Hence, we can choose the sector for which a CNOT gate is applied simply by choosing whether the constituent Hadamard gates operate in the metastable sector (red, via pulses with $\Omega_m$) or the optical clock sector (blue, via pulses with $\Omega_c$). This versatility is afforded by the more general `omg' architecture~\cite{Chen2022,Lis2023}.

Before describing the protocol, we note that our state will generally reside within the \{gab\} qutrit Hilbert space. We must therefore devise a readout protocol that can differentiate these three states, which will require two rounds. Figure~\ref{Figure2}(b) shows this readout operation in which fluorescence readout is applied to one or both of the nuclear spin $^1$S$_0$ ground states, leaving the metastable manifold dark and unaffected. By placing a $\pi$-pulse on the clock transition between the two readout operations~\cite{Huie2023,Norcia2023,Jia2024}, we can uniquely identify the state of the qutrit (assuming the atom has not been lost -- a possibility that could be detected with a third measurement round).

The protocol is schematically depicted in Fig.~\ref{Figure2}(c), in which we have three atomic processor nodes that share Bell pairs as an initial resource. See below for further detail. First, we construct a three-qubit W-state 
\begin{equation}
|W\rangle=\frac{1}{\sqrt{3}}\big(|a,g,g\rangle+|g,a,g\rangle+|g,g,a\rangle\big)
\end{equation}
in Node 1, which resides in the \{ga\} sector. We then teleport two of these qubits to Nodes 2 and 3. From here, we could either begin the distributed clock interrogation protocol, or we could amplify the effective qubit energy by $N\times$ by spreading the state of each qubit into an $N$-atom GHZ state in each node. In either case, our atoms now reside in the full \{gab\} qutrit space. We probe the differential proper time for a duration $T$ (as measured in Node 1), and then we prepare for measurement. First, we unbuild the GHZ state if we used it, and then we teleport the qutrits in Nodes 2 and 3 back to Node 1. Finally, we perform a nonlocal measurement that is blind to the identity of our three atoms, yet results in a periodic loss and revival of the visibility of interferometric ``self-interference" between all three clocks as a signature of the interplay of quantum dynamics and general relativistic proper time evolution. 

We assume that the `master clock laser' is in Node 1 and is referenced to a separate atomic clock to ensure long-term phase stability with respect to atoms. The `master clock' also defines the `wall time' $T$ for each node. The stability of this laser is transferred to Nodes 2 and 3 via standard phase-stabilized fiber link techniques~\cite{Ma1994,Pizzocaro2021,Gozzard2022}, where it is used to `lock' the local oscillators in Nodes 2 and 3, ensuring a common frequency reference in all nodes. This approach is ubiquitous in classical clock networks~\cite{Ludlow2015,Wcislo2018,Pizzocaro2021,Gozzard2022}.

\begin{figure*}[t!]
    \centering
    \includegraphics[width=0.8\textwidth]{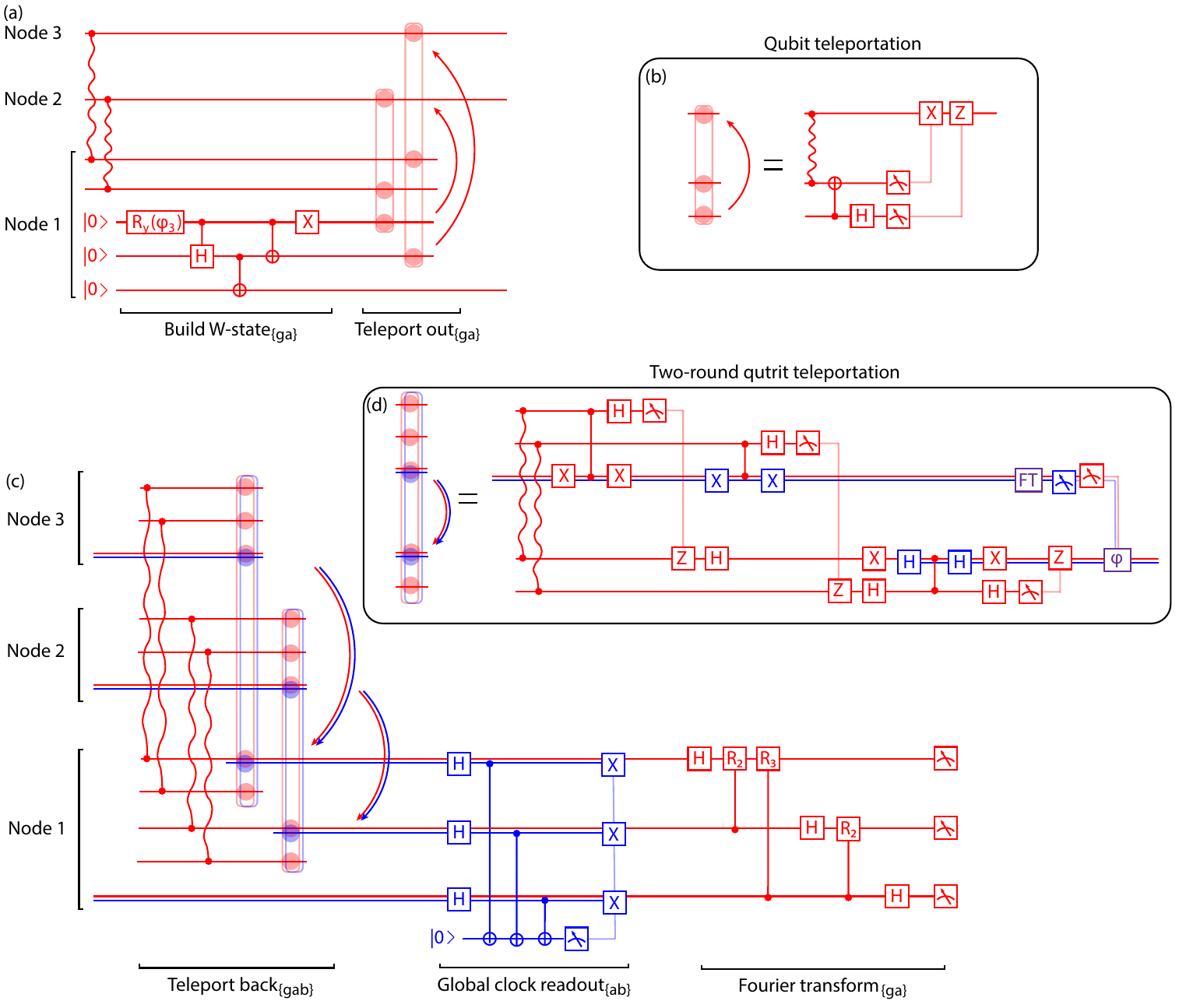}
    \caption{
        \textbf{Creating, distributing, and detecting W-states.} (a) To initialize a distributed W-state, we begin by distributing a Bell pair between Nodes 1 and 2 and a Bell pair between Nodes 1 and 3. Then we produce a 3-atom W-state in Node 1 in the $\{ga\}$-sector, where $\phi_3=2\text{arccos}(1/\sqrt{3})$. We teleport two qubits in the W-state, one each to Nodes 2 and 3 using a standard qubit teleportation protocol shown in (b) and based on auxiliary entanglement. Then, a GHZ state may be built and unbuilt within each node following the protocol in Fig.~\ref{Figure5}, sandwiched around an interrogation of GR through proper time measurements as above. (c) This leaves us in a distributed W-state in the $\{gab\}$ qutrit space. We begin by teleporting the qutrits in Nodes 2 and 3 back to Node 1 using two Bell pairs each of auxiliary entanglement. The two-round qutrit teleportation protocol is shown in (d). Once the qutrit W-state has been returned to Node 1, an ancilla-based non-local probe of whether the clock qubit is in $|a\rangle$ or $|b\rangle$ (called a `global clock readout') is performed, followed by a conditional global $X$ operation on the $\{ab\}$-sector. Then a Fourier transform is applied for non-local readout of the $\{ga\}$-sector.
        \label{Figure3}
    }
\end{figure*}

\section{Signature of quantum interference in the presence of spacetime curvature}
We now discuss how to create, distribute, and detect W-states in order to obtain a quantum interference signal of proper times that contains signatures of curved spacetime. We begin by building the distributed W-state locally in Node 1, and then teleport two of its constituent qubits to Nodes 2 and 3. This circuit is shown in Fig.~\ref{Figure3}(a), where the standard qubit teleportation protocol based on auxiliary entanglement is shown in Fig.~\ref{Figure3}(b). The W-state resides in the $\{ga\}$-sector. After building the non-local W-state (neglecting the GHZ state for now), we start the non-local clock by applying a $\pi/2$ rotation in the $\{ab\}$ subspace similar to the protocol of Ref.~\cite{borregaard2024testing}. The resulting state after a certain free evolution time is:
\begin{eqnarray} \label{eq:evolved}
|W_T\rangle&=&\frac{1}{\sqrt{6}}\big((|a\rangle+e^{-i\theta_1}|b\rangle)|g\rangle|g\rangle\nonumber \\
&+&e^{-i\phi_1}|g\rangle (|a\rangle+e^{-i\theta_2}|b\rangle)|g\rangle\nonumber \\
&+&e^{-i\phi_2}|g\rangle|g\rangle(|a\rangle+e^{-i\theta_3}|b\rangle)\big),
\end{eqnarray}
where $\theta_j=(E_b-E_a)\tau_j/\hbar=\Delta E \tau_j/\hbar$, $\phi_j=(E_g-E_a)(\tau_{1}-\tau_{j+1})/\hbar$, and $\tau_j$ is the proper time at the $j$'th node.  
The states $|c(\tau_i)\rangle \equiv \frac{1}{\sqrt{2}} \left( |a\rangle+e^{-i\theta_1}|b\rangle \right)$ are the local clock states that record the proper time $\tau_i$ that depends on the position in the gravitational field. We must now perform a non-local measurement of our W-state. We first teleport our qutrits back to Node 1 such that all subsequent operations are local [see Fig~\ref{Figure3}(c)]. Naturally, teleporting a qutrit is more advanced than teleporting a qubit, and two auxiliary Bell pairs are required for each qutrit. As shown in Fig.~\ref{Figure3}(d), the qutrit teleportation can be performed in two rounds. A phase correction on the teleported qutrit state conditional on the teleportation measurement outcomes may be required. Then, the final step involves a quantum Fourier transform of our initial qutrit followed by measurement. 

Once all the qutrits are in Node 1, we use an ancilla qubit to measure the state of the clock qubit without revealing its location. First, we perform a Hadamard gate in the $\{ab\}$ subspace on each system. After this, we perform a series of CNOT gates in the $\{ab\}$-sector with each qutrit in the W-state serving as a control and the ancilla serving as the target such that a subsequent measurement of the ancilla qubit will project the state into either the $\{ga\}$ or $\{gb\}$ subspace. 

We then apply a global $X$ pulse in the $\{ab\}$ sector conditional on the ancilla measurement outcome to ensure that the post measurement state is entirely in the $\{ga\}$-subspace and will be of the form
\begin{eqnarray}
\ket{\psi_{\pm}}&=&\frac{1}{6}\big((1\pm e^{-i\theta_1})\ket{a,g,g}+e^{-i\phi_1}(1\pm e^{-i\theta_2})\ket{g,a,g} \nonumber \\
&&+e^{-i\phi_2}(1\pm e^{-i\theta_3})\ket{g,g,a}\big),
\end{eqnarray}
with $\ket{\psi_+}$ ($\ket{\psi_-}$) corresponding to measuring the auxiliary atom in state $\ket{a}$ ($\ket{b}$). The norm $|\braket{\psi_+}{\psi_+}|^2$ ($|\braket{\psi_-}{\psi_-}|^2$) corresponds to the probability of measuring $\ket{a}$ ($\ket{b}$). To interfere the three clock locations, we subsequently perform a measurement in the Fourier basis $\{\ket{x}\}$ where
\begin{eqnarray} \label{eq:Fourier}
\ket{x}=\frac{1}{\sqrt{3}}\left(\ket{a,g,g}+e^{iw_x}\ket{g,a,g}+e^{2iw_x}\ket{g,g,a}\right),
\end{eqnarray}
with $w_x=2\pi x/3$ and $x\in\{0,1,2\}$. Defining the projectors, $\Pi_x=\ket{x}\bra{x}$, we see that
\begin{eqnarray} \label{eq:wpovm}
\avg{\Pi_x}&=&\bra{\psi_+}\Pi_x\ket{\psi_+}+\bra{\psi_-}\Pi_x\ket{\psi_-} \nonumber \\
&=&\frac{1}{9}\big(3+\cos(w_x-\phi_2+\phi_1)+\cos(\theta_1-\theta_2-w_x-\phi_1) \nonumber \\
&&+\cos(2w_x-\phi_2)+\cos(\theta_1-\theta_3-2w_x-\phi_2) \nonumber \\
&&+\cos(w_x-\phi_1)+\cos(\theta_3-\theta_2+w_x+\phi_2-\phi_1) \big)\ \nonumber \\
&=&\frac{1}{3}+\frac{2}{9}\Big(|\braket{c(\tau_1)}{c(\tau_2)}|\cos(\lambda_{12}-w_x-\phi_1) \nonumber \\ \label{eq:observable}
&&+|\braket{c(\tau_1)}{c(\tau_3)}|\cos(\lambda_{13}-2w_x-\phi_2) \nonumber \\
&&+|\braket{c(\tau_2}{c(\tau_3)}|\cos(\lambda_{23}-w_x-\phi_2+\phi_1),
\end{eqnarray}
where we have defined $\ket{c(\tau_i)}=\frac{1}{\sqrt{2}}(\ket{a}+e^{-i\Delta E\tau_i/\hbar}\ket{b})$, and $\braket{c(\tau_i)}{c(\tau_j)}=|\braket{c(\tau_i)}{c(\tau_j)}|e^{-i\lambda_{ij}}$. From Eq.~(\ref{eq:wpovm}), we see that the measured projectors exhibit oscillating terms due to the gravitational time-dilation between all pairwise combinations of the three clock locations.   

\begin{figure}[t!] 
    \centering 
\begin{subfigure}[t]{0.5\textwidth} 
        \centering
        \includegraphics[width=1.0\textwidth]{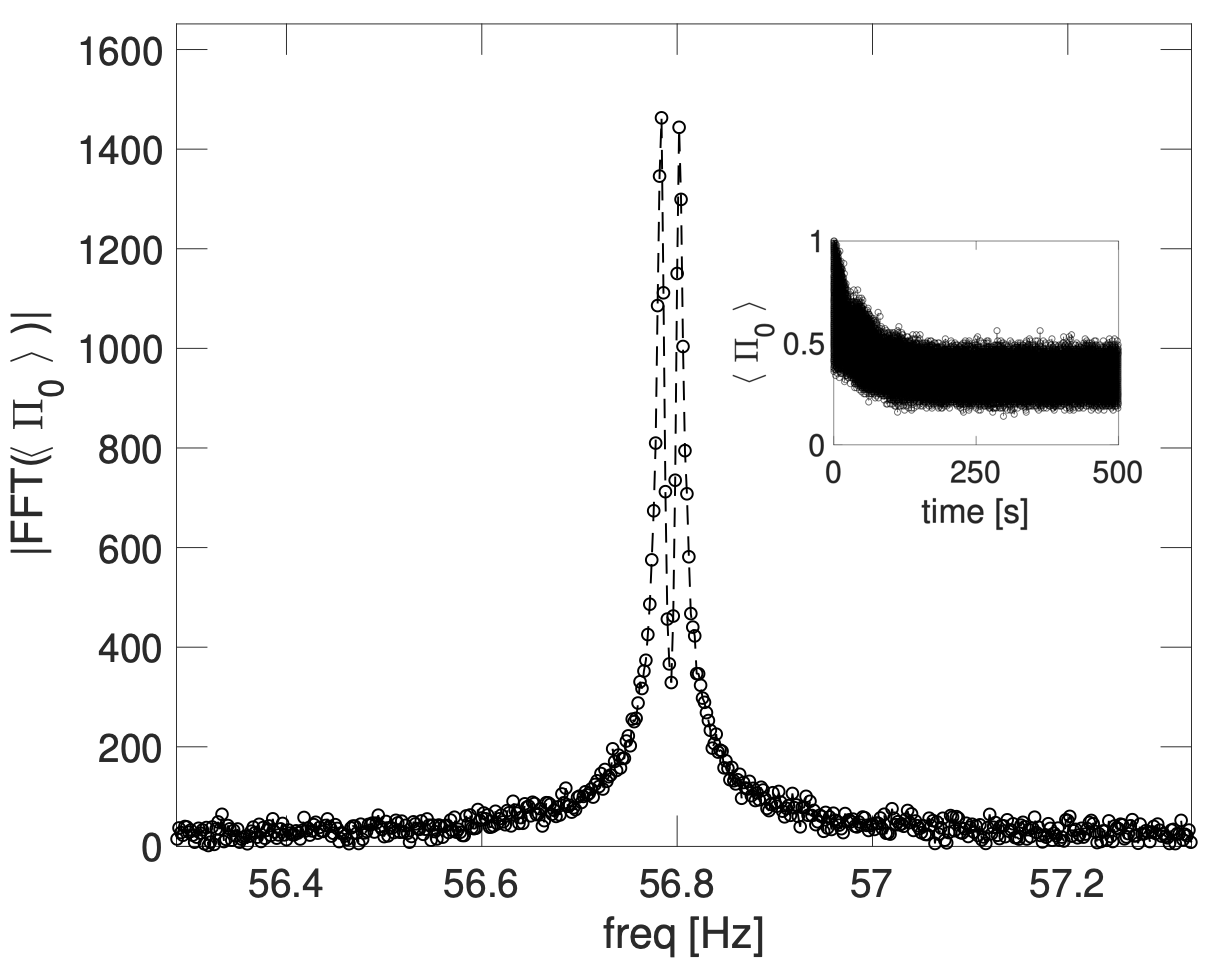}
    \end{subfigure} \\
    \begin{subfigure}[t]{0.5\textwidth}
        \centering
        \includegraphics[width=1.0\textwidth]{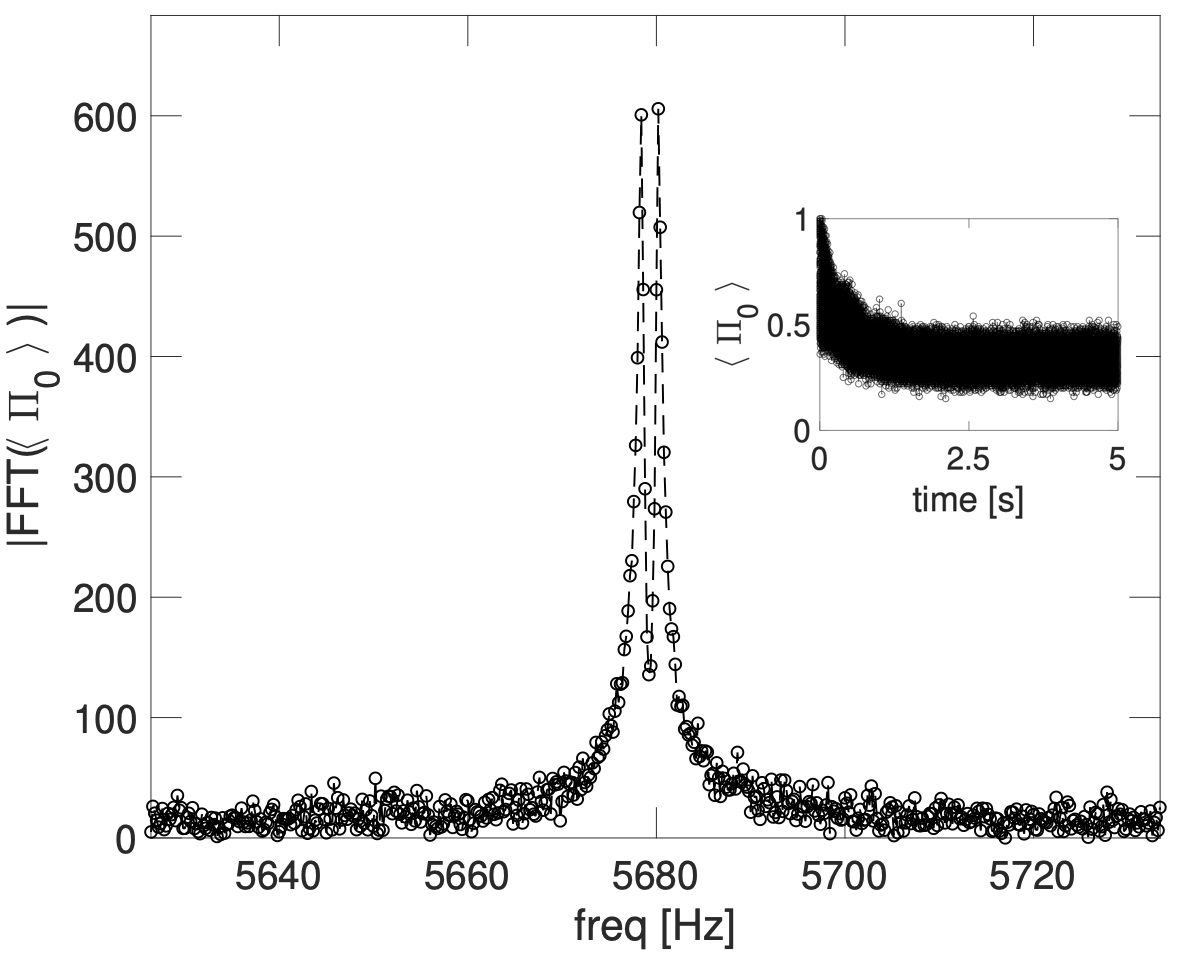}
    \end{subfigure}
     \caption{(top) Fast Fourier Transform of our non-local clock observable of interest in eq. \eqref{eq:wpovm}, $\langle\Pi_0\rangle$, assuming three atoms positioned at Earth's surface, 1 km, and 2 km above the surface, respectively. The signal is sampled at a rate of 0.5 kHz for a total evolution time of $T=500$ sec. Each data point is calculated as the mean of 100 samples and we have assumed that the atoms are subject to dephasing noise corresponding to $T_2=50$ sec. The line splitting arises due to spacetime curvature. The inset shows the sampled signal as a function of time. (bottom) Same plot but for three GHZ-super atoms with $N=100$. The sampling rate is 10 kHz and we have assumed an effective GHZ dephasing time of 0.5 sec.}\label{Figure4}
\end{figure}

The observable captures interference of the three proper time evolutions at their respective locations, which provides a genuine test of quantum mechanics in curved spacetime. As an example, we consider a scenario where the three locations are placed at distances $d_j=jd$ ($j\in{0,1,2}$) from Earth's surface. For $^{171}$Yb, we have that $E_g\approx E_a$ while $(E_a-E_b)/h=-\Delta E/\hbar\approx 520$ THz. The measured signal $\avg{\Pi_x}$ for any choice of $x$ would exhibit oscillations with the three frequencies
\begin{eqnarray}
\omega_{12}&=&\frac{\partial}{\partial t}(\theta_1-\theta_2)=\frac{\Delta E \cdot GM}{\hbar c^2} \left(\frac{1}{R}-\frac{1}{R+d}\right) \quad \\
\omega_{23}&=&\frac{\partial}{\partial t}(\theta_2-\theta_3)=\frac{\Delta E \cdot G M}{\hbar c^2} \left(\frac{1}{R+d}-\frac{1}{R+2d}\right)\quad\\
\omega_{13}&=&\frac{\partial}{\partial t}(\theta_1-\theta_3)=\frac{\Delta E \cdot G M}{\hbar c^2} \left(\frac{1}{R}-\frac{1}{R+2d}\right).
\end{eqnarray}
To first order, we see that $\omega_{12}\approx\omega_{23}$ but the effect of curvature is visible in the higher order terms since
\begin{equation}\label{eq:frequency}
\Delta\omega=\omega_{12}-\omega_{23}\approx2\frac{\Delta E \cdot G M}{\hbar c^2}\frac{d^2}{R^3},
\end{equation}
to leading order. The resulting post-Newtonian frequency shift thus shows simultaneous effects of both curved spacetime and quantum interference. To resolve this frequency, we need to let the system evolve for a total `wall time' of 
\begin{equation}
T\approx\frac{1}{\Delta\omega}=\frac{\hbar R^3 c^2 }{2d^2\Delta E \cdot G M}. 
\end{equation}
In figure~\ref{Figure4}, we plot $\langle\Pi_0\rangle$ from three single atom clocks for $d=1$ km. It is seen that for a single atom coherence time of $T_2=50$ sec, it is possible to resolve $\Delta \omega \sim 0.02$ Hz for a total evolution time of $T=500$ sec. Although obtaining such coherence times is challenging, it is within reach of current hardware~\cite{Young2020,Bothwell2022}. Moreover, ``spectator" ancilla qubits could be employed to track and compensate global magnetic field noise~\cite{Singh2022}, which is the leading source of decoherence for nuclear Zeeman states.

\begin{figure*}[t!]
    \centering
    \includegraphics[width=0.85\textwidth]{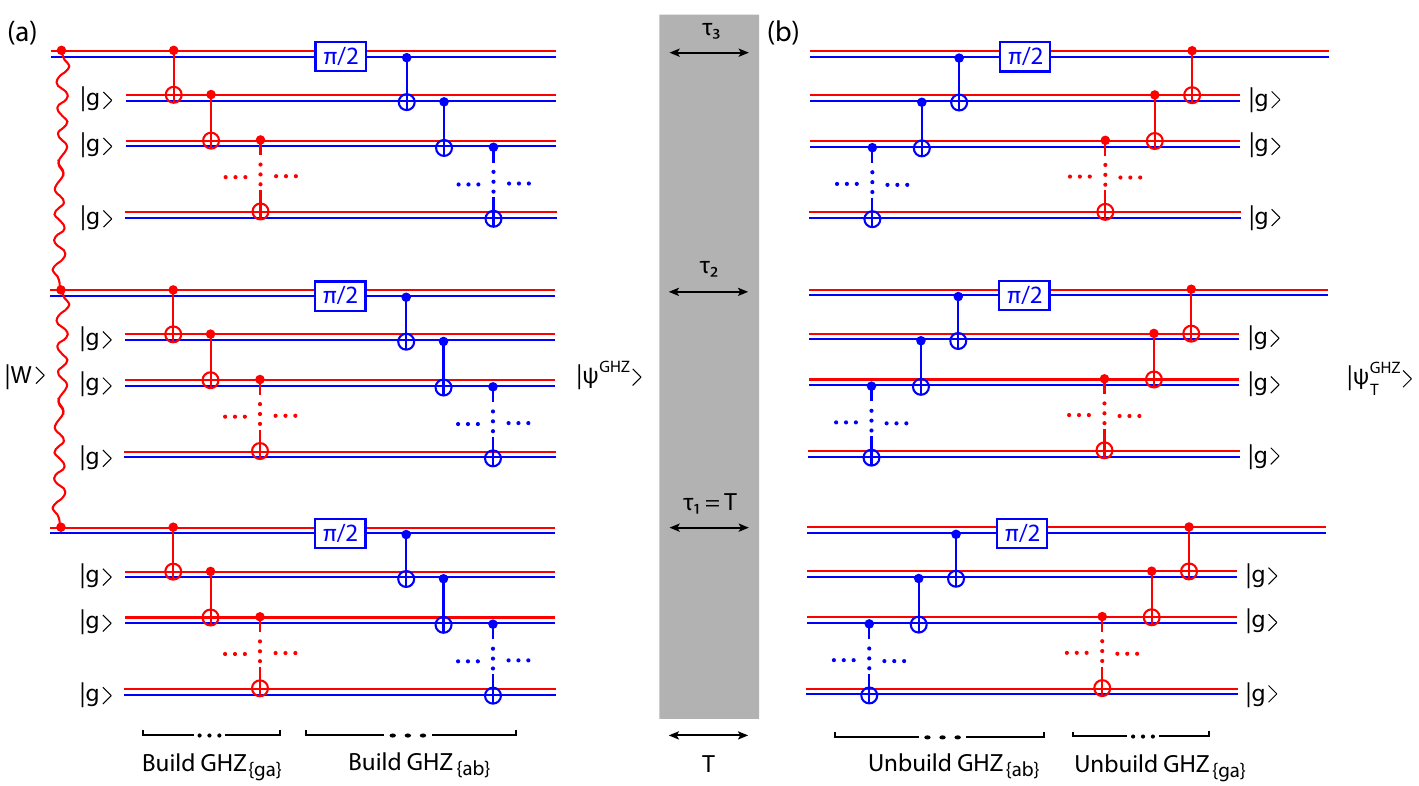}
    \caption{
        \textbf{Building and unbuilding distributed GHZ super-atoms.} (a) We begin with the distributed W-state $|W\rangle$ from Fig.~\ref{Figure3} as well as $N-1$ additional qubits initialized in $|g\rangle$. We then build a GHZ state in the \{ga\}-sector in each module with cascaded CNOT gates. The resulting state is now a version of $|W\rangle$ in which every single-atom state is raised to the $\otimes N$. Then, we apply a $\pi/2$-pulse to the clock transition on the ``science" qubits, and use them to grow a GHZ state in the \{ab\}-sector using cascaded CNOT gates that operate in that sector (blue; $|a\rangle\equiv|1\rangle$ and $|b\rangle\equiv|0\rangle$). Then, the clock Ramsey interferometer now works in the same way as the single-atom case except that the effective phase accrual due to the difference in proper time is $N\times$ larger. This can be equivalently viewed as a single super-atom with a $N\times$ larger clock transition energy. Note that this distributed GHZ state spans the \{gab\} qutrit space, like the single-atom case. (b) To measure this state, we must unbuild the GHZ states such that we can perform the non-local measurement. We unbuild in both the \{ab\} and \{ga\} sectors with cascaded CNOT gates, essentially running time backwards compared to the build operations. The additional $N-1$ qubits in each module return to $|g\rangle$, and the ``science" W-state is in a phase-multiplied version of the state $|W_T\rangle$. We can now apply the protocol from Fig.~\ref{Figure3} for non-local measurement. 
        \label{Figure5}
    }
\end{figure*}

\section{Entanglement-enhanced sensing using GHZ super-atoms}
One challenge with the assumed $T_2=50$ and $T=500$ sec for this measurement is the presence of technical noise sources that can have a nonlinear dependence on the interrogation time. These can include thermal effects in stabilization electronics for optical potentials and magnetic fields, and even heating of the atom in an optical trap due to intensity noise. Moreover, we propose to perform this experiment in a terrestrial network which will require one node to be at the top of a tall building. Therefore, we must be careful to obviate the effects of natural, low-frequency oscillations. Because of these effects and others, the `wall clock' time for this experiment is an important consideration, and methods to boost the measurement bandwidth are worth considering. 

It is possible to boost $\Delta\omega$ without increasing the separation between the atomic nodes. We can employ $N$-atom GHZ states in each system to effectively create a ``super-atom" with a $N\times$ larger clock transition energy. The protocol for creating and performing the Ramsey  interrogation with the GHZ state is shown in Fig.~\ref{Figure5}(a). We again start with the W state
\begin{equation}
|W\rangle=\frac{1}{\sqrt{3}}\big(|a,g,g\rangle+|g,a,g\rangle+|g,g,a\rangle\big)
\end{equation}
and then we build a GHZ state in each module through the use of cascaded CNOT gates with the other $N-1$ qubits that were initially in $|g\rangle$. This creates the state
\begin{eqnarray}
|\psi\rangle&=&\frac{1}{\sqrt{6}}(|a\rangle^{\otimes N}|g\rangle^{\otimes N}|g\rangle^{\otimes N}+|g\rangle^{\otimes N}|a\rangle^{\otimes N}|g\rangle^{\otimes N}\nonumber \\
&+&|g\rangle^{\otimes N}|g\rangle^{\otimes N}|a\rangle^{\otimes N}).
\end{eqnarray}
The local GHZ states resides in the \{ga\}-sector, and we must similarly build GHZ states in the \{ab\} sector to start the non-local clock. We apply a $\pi/2$-pulse on the clock transition of the ``science" atoms, and then perform the cascaded CNOT gates, now in the \{ab\}-sector [blue in Fig.~\ref{Figure5}(a)]. 

We are now in the state
\begin{eqnarray}
|\psi^\text{GHZ}\rangle&=&\frac{1}{\sqrt{6}}\bigg(\big(|a\rangle^{\otimes N}+|b\rangle^{\otimes N}\big)|g\rangle^{\otimes N}|g\rangle^{\otimes N} \nonumber \\
&+&|g\rangle^{\otimes N}\big(|a\rangle^{\otimes N}+|b\rangle^{\otimes N}\big)|g\rangle^{\otimes N} \nonumber \\
&+&|g\rangle^{\otimes N}|g\rangle^{\otimes N}\big(|a\rangle^{\otimes N}+|b\rangle^{\otimes N}\big)\bigg).
\end{eqnarray}
After the evolution under the presence of differential proper time, we are in the state
\begin{eqnarray}
|\psi^\text{GHZ}_T\rangle\!\!&=&\!\!\frac{1}{\sqrt{6}}\bigg(\big(|a\rangle^{\otimes N}+e^{-iN\theta_1}|b\rangle^{\otimes N}\big)|g\rangle^{\otimes N}|g\rangle^{\otimes N} \nonumber \\
\!\!&+&\!\!e^{-iN\phi_1}|g\rangle^{\otimes N}\big(|a\rangle^{\otimes N}+e^{-iN\theta_2}|b\rangle^{\otimes N}\big)|g\rangle^{\otimes N} \nonumber \\
\!\!&+&\!\!e^{-iN\phi_2}|g\rangle^{\otimes N}|g\rangle^{\otimes N}\big(|a\rangle^{\otimes N}+e^{-iN\theta_3}|b\rangle^{\otimes N}\big)\bigg). \qquad
\end{eqnarray}

To interfere the clock state of the different locations, the measurement protocol must be compatible with the non-local measurement circuits shown in Figs.~\ref{Figure2} and~\ref{Figure3}. As such, we want to remove the extra $N-1$ qubits used in the GHZ super-atom in each system. We thus unbuild the GHZ state, as shown in Fig.~\ref{Figure5}(b). The protocol again uses cascaded CNOT gates in reverse order, essentially to ``undo" the entanglement from the cascaded CNOT gates in the building process. We begin in the \{ab\}-sector by applying the cascaded CNOT gates and the $\pi/2$-pulses on the clock transition of the ``science" atoms. We then unbuild the GHZ state in the \{ga\}-sector. This leaves us in the state shown in Eq.~(\ref{eq:evolved}) except that all phases have been multiplied by a factor of $N$. The final non-local measurement can now be performed as previously described for the simple W-state.  

Using GHZ super atoms, we effectively increase the interrogation bandwidth of the tests by a factor of $N$ allowing for a much shorter interrogation time. In figure~\ref{Figure4}(b), we plot $\langle\Pi_0\rangle$ from three GHZ super-atoms at separations of $d=1$ km. Each super-atom consists of $N=100$ atoms and we have assumed an effective dephasing time of the GHZ state super-atoms of $0.5$ s ($N\times$ faster). It is possible to resolve $\Delta \omega \sim 2 $ Hz for a total evolution time of $5$ s.      

\section{Testing quantum theory on curved spacetime}
The central science goal of this proposal is to observe how quantum dynamics and general relativistic curved spacetime intertwine, in a single observable. If one of those theories are modified in this regime, the outcome would be different than predicted here. The key observable is given in Eq.~\eqref{eq:observable}. It shows the interference of the clock state overlaps $\braket{c(\tau_x)}{c(\tau_y)}$ which are different from unity due to the different general relativistic proper time evolutions $U(\tau_x)$ at the locations of the clocks. The observable involves not just these proper time evolutions as in regular clock measurements, but also their quantum interference. Spacetime curvature manifests itself in the scaling of the proper times: they are no longer changing linearly with height $h$ as $\Delta \tau = g h t/c^2$, which would also be the case in special relativistic acceleration. The observable thus cannot be mapped from gravity to special relativity according to the equivalence principle, as curved spacetime induces a nonlinear spatial scaling between the different proper times. This manifests itself in the line-splitting in frequency space of the observables $\langle \Pi_x \rangle$, as in figure \ref{Figure4}. The test also clearly goes beyond any Newtonian curvature effect, as the observable is due to the general relativistic proper time. This is in contrast to already performed superposition experiments with atomic fountains that are sensitive to spacetime curvature in the Newtonian limit \cite{asenbaum2017phase}.

In what way could the outcomes differ from expected physics? When gravity and quantum theory intertwine, deviations from regular quantum physics might emerge \cite{penrose1996gravity, bekenstein2014can, ralph2014entanglement, rydving2021gedanken}, and no empirical evidence of the validity of quantum theory in this regime exists so far. Our protocol enables several critical tests of this domain. Firstly, as outlined in Ref.~\cite{borregaard2024testing}, such an experiment with entangled clocks can be used to directly test the unitarity and linearity of quantum theory on curved spacetime. To this end, one can compare two different experimental protocols: one using entangled clocks as proposed above, and another one with initially separable clocks and the same non-local measurement operators $\Pi_x$. The latter should also produce the exact same \textit{conditional} outcome as in Eq.~\eqref{eq:observable}, but with finite success probability. This is because in normal linear and unitary quantum evolution, the non-local contribution is one of many amplitudes that contribute to the same outcome, even for initial product states. It can be easily seen when writing the state in terms of Bell-states, which constitute a basis. Explicitly, the protocol for the conditional experiment would be the following: one first prepares the three clocks in the product state $\ket{\Psi(0)} = \Pi_{i=1}^{\otimes 3} \left(\frac{1}{\sqrt{3}}\left(\ket{g} + \sqrt{2}\ket{c(0)} \right)  \right)$ where $\ket{c(0)} = \frac{1}{\sqrt{2}} \left(\ket{a} + \ket{b} \right)$. The states evolve to $\ket{\Psi} = \frac{1}{\sqrt{27}} \Pi_{i=1}^{\otimes 3} \left(e^{-i E_g \tau_i} \ket{g} + \sqrt{2} e^{-i E_b \tau_i}\ket{c(\tau_i)} \right)   $. This state, while being a product state, has also the amplitude $\sqrt{\frac{2}{9}}e^{- i \varphi}\ket{W_T}$ from Eq. \eqref{eq:evolved}, on which we conditionally project yielding the same outcome as in Eq.~\eqref{eq:observable}. Thus the comparison of outcomes between initially entangled and product states would provide a direct test of unitarity and linearity of quantum theory even when the dynamics must include general relativistic corrections. The only other performed test of unitarity in the presence of spacetime curvature, to our knowledge, focused on the propagation of photons in Earth's gravitational field to test for anomalous decoherence~\cite{xu2019satellite}. Our proposed test with time dilation of entangled clocks would broaden the scope to test the validity of quantum theory for genuine general relativistic observables.

Secondly, our protocol explores the quantum dynamics with gravitational corrections both in the general relativistic limit and beyond the homogeneous limit. This opens new opportunities to test gravity in quantum theory as one can see from a semi-classical phase-space picture: as long as the potentials are at most quadratic in position, the dynamics if represented by a Wigner quasi-probability distribution is indistinguishable from a classical Liouville equation. Thus no quantum contributions arise from the dynamics in this limit \cite{marchese2024newton}. However, this mapping breaks down when the dynamics becomes non-Gaussian. Pushing into this limit enables tests of how curved spacetime may affect the quantum dynamics in anomalous ways \cite{plavala2025probing}. Our protocol operates in this limit and the proper time observables enable the exploration of such possibly anomalous dynamics in the post-Newtonian regime. To probe genuine quantum dynamics in the Wigner evolution amounts to observing time-dilation differences due to non-linear potential differences beyond the second order, when equation \eqref{eq:frequency} gets the additional contribution $\Delta \omega_3  \approx 5\frac{\Delta E \cdot G M}{\hbar c^2}\frac{d^3}{R^4}$.

And thirdly, our test also provides a novel exploration of one of the most fundamental principles of quantum theory: the Born rule. One of its core consequences is intereference of amplitudes, where the probability of superposed configurations $P_{12}$ leads to interference $I_{12}$ in addition to the individual probabilities, $P_{12} = P_1 + P_2 + I_{12}$. Higher-order interference, however, does not arise \cite{sorkin1994quantum}, as the Born rule results only in pairwise interference terms. Our observable in \eqref{eq:observable} is constructed from three-path-interference of the three clocks, as indicative by the projection measurement in the Fourier basis, Eq.~\eqref{eq:Fourier}. One can combine such measurements with observables constructed only with two clocks interfering, or even just individual clocks. One can construct the second-order interference quantity 
\begin{equation} \label{eq:Born}
    I_{123} = P_{123} - P_{12} - P_{13} - P_{23} + P_1 + P_2 +P_3 \, ,
\end{equation}
where $P_{ijk}$ correspond to the probabilities of obtaining the observable with the clocks in the respective positions. In regular quantum theory, this quantity exactly vanishes ($I_{123}=0$) due to Born's rule. However, higher-order interferences that go beyond the Born rule would show in a non-zero value. This has been used to experimentally test the Born rule \cite{sinha2010ruling, park2012three}. It has also been suggested that quantum gravity might result in modifications to the Born rule \cite{valentini2023beyond}, with early explorations by Sorkin \cite{sorkin1994quantum}. Our protocol here can easily implement the measurement \eqref{eq:Born} in analogy to \cite{park2012three}. We measure directly the 3-part interference $P_{ijk}$, which is what the outcome \eqref{eq:observable} yields. One can now extend the protocol to also measure only two-clock interefences, and even single clock measurements as required by eq \eqref{eq:Born}. To compare them one has to make sure the amplitudes remain unchanged from $1/\sqrt{3}$. This simply requires an additional ancillary state as in Ref. \cite{park2012three}, which again is straightforward to implement in our protocol. Thus the setup we propose can easily be used to probe the Born rule and can therefore provide the first empirical test of one of the most fundamental principles of quantum theory under the influence of general relativity.

\section{Concluding discussion}

Using quantum networking techniques, we have proposed a protocol to probe general relativistic time dilation due to the curvature of Earth's gravitational potential via three entangled atomic systems with $\sim$km-scale separation. Our proposal combines many state-of-the-art techniques in atomic quantum science ranging from quantum networking~\cite{Covey2023} to quantum algorithms~\cite{Graham2022,Bluvstein2022} to quantum metrology~\cite{Finkelstein2024,Cao2024}. Further, we utilize qutrit code spaces composed of nuclear and optical clock degrees of freedom, with inspiration from recent work on qudit~\cite{Ringbauer2022} and multi-qubit encodings~\cite{Jia2024} in atomic systems. We have mostly focused on neutral ytterbium-171 atoms, but our protocols could equally be applied to trapped ion systems. Recently, remote entanglement between two ions with a fidelity of 0.97 has been demonstrated by using a time bin encoding scheme~\cite{Saha2024}, and entanglement between the nuclear spin and a telecom-band photon in neutral ytterbium-171 atoms has recently been demonstrated with a fidelity of 0.97~\cite{Li2025}. Given the demonstrated coherence times of nuclear spin qubits ($\sim$7 seconds)~\cite{Lis2023} and the demonstrated half-minute-scale atom-atom coherence of optical clock qubits~\cite{Young2020,Bothwell2022} as well as the demonstrated Rydberg-mediated gate fidelities of $\sim$0.995 across several different qubit encodings~\cite{Evered2023,Finkelstein2024,Peper2024,Radnaev2024,Muniz2024}, we anticipate that it may be feasible to realize $\sim$100-atom GHZ states in each node with fidelity above $\sim$0.5.

The proposed setup enables unique tests of the interplay of quantum theory and general relativity -- where no such tests have yet been performed. We showed that it can test whether some of the most fundamental tenants of quantum theory survive in the presence of spacetime curvature, focusing on the interference of general relativistic proper time evolutions. Our proposed tests of the unitarity, linearity and the probabilistic Born rule of quantum theory enable the first exploration of the interplay of the foundations of quantum theory and general relativity in a realistic experiment with quantum networks, opening the door for unique new ground-based and space-based tests of quantum physics and general relativity.

\section*{Acknowledgments}
J.P.C. acknowledges funding from the NSF QLCI for Hybrid Quantum Architectures and Networks (NSF award 2016136); the NSF PHY Division (NSF awards 2112663 and 2339487); the NSF Quantum Interconnects Challenge for Transformational Advances in Quantum Systems (NSF award 2137642); the ONR Young Investigator Program (ONR award N00014-22-1-2311); the AFOSR Young Investigator Program (AFOSR award FA9550-23-1-0059); and the U.S. Department of Energy, Office of Science, National Quantum Information Science Research Centers. I.P. acknowledges support by the NSF under award No 2239498, NASA under award No 80NSSC25K7051 and the Sloan Foundation under award No G-2023-21102. J.B. acknowledges support from The AWS Quantum Discovery Fund at the Harvard Quantum Initiative. \\

\bibliography{library}

\begin{thebibliography}{78}%
\makeatletter
\providecommand \@ifxundefined [1]{%
 \@ifx{#1\undefined}
}%
\providecommand \@ifnum [1]{%
 \ifnum #1\expandafter \@firstoftwo
 \else \expandafter \@secondoftwo
 \fi
}%
\providecommand \@ifx [1]{%
 \ifx #1\expandafter \@firstoftwo
 \else \expandafter \@secondoftwo
 \fi
}%
\providecommand \natexlab [1]{#1}%
\providecommand \enquote  [1]{``#1''}%
\providecommand \bibnamefont  [1]{#1}%
\providecommand \bibfnamefont [1]{#1}%
\providecommand \citenamefont [1]{#1}%
\providecommand \href@noop [0]{\@secondoftwo}%
\providecommand \href [0]{\begingroup \@sanitize@url \@href}%
\providecommand \@href[1]{\@@startlink{#1}\@@href}%
\providecommand \@@href[1]{\endgroup#1\@@endlink}%
\providecommand \@sanitize@url [0]{\catcode `\\12\catcode `\$12\catcode
  `\&12\catcode `\#12\catcode `\^12\catcode `\_12\catcode `\%12\relax}%
\providecommand \@@startlink[1]{}%
\providecommand \@@endlink[0]{}%
\providecommand \url  [0]{\begingroup\@sanitize@url \@url }%
\providecommand \@url [1]{\endgroup\@href {#1}{\urlprefix }}%
\providecommand \urlprefix  [0]{URL }%
\providecommand \Eprint [0]{\href }%
\providecommand \doibase [0]{https://doi.org/}%
\providecommand \selectlanguage [0]{\@gobble}%
\providecommand \bibinfo  [0]{\@secondoftwo}%
\providecommand \bibfield  [0]{\@secondoftwo}%
\providecommand \translation [1]{[#1]}%
\providecommand \BibitemOpen [0]{}%
\providecommand \bibitemStop [0]{}%
\providecommand \bibitemNoStop [0]{.\EOS\space}%
\providecommand \EOS [0]{\spacefactor3000\relax}%
\providecommand \BibitemShut  [1]{\csname bibitem#1\endcsname}%
\let\auto@bib@innerbib\@empty
%</preamble>
\bibitem [{\citenamefont {Oriti}(2009)}]{oriti2009approaches}%
  \BibitemOpen
  \bibfield  {author} {\bibinfo {author} {\bibfnamefont {D.}~\bibnamefont
  {Oriti}},\ }\href@noop {} {\emph {\bibinfo {title} {Approaches to quantum
  gravity: Toward a new understanding of space, time and matter}}}\ (\bibinfo
  {publisher} {Cambridge University Press},\ \bibinfo {year}
  {2009})\BibitemShut {NoStop}%
\bibitem [{\citenamefont {Fulling}(1989)}]{fulling1989aspects}%
  \BibitemOpen
  \bibfield  {author} {\bibinfo {author} {\bibfnamefont {S.~A.}\ \bibnamefont
  {Fulling}},\ }\href@noop {} {\emph {\bibinfo {title} {Aspects of quantum
  field theory in curved spacetime}}},\ \bibinfo {number} {17}\ (\bibinfo
  {publisher} {Cambridge university press},\ \bibinfo {year}
  {1989})\BibitemShut {NoStop}%
\bibitem [{\citenamefont {Carney}\ \emph {et~al.}(2019)\citenamefont {Carney},
  \citenamefont {Stamp},\ and\ \citenamefont {Taylor}}]{carney2019tabletop}%
  \BibitemOpen
  \bibfield  {author} {\bibinfo {author} {\bibfnamefont {D.}~\bibnamefont
  {Carney}}, \bibinfo {author} {\bibfnamefont {P.~C.}\ \bibnamefont {Stamp}},\
  and\ \bibinfo {author} {\bibfnamefont {J.~M.}\ \bibnamefont {Taylor}},\
  }\bibfield  {title} {\bibinfo {title} {Tabletop experiments for quantum
  gravity: a user’s manual},\ }\href
  {https://doi.org/10.1088/1361-6382/aaf9ca} {\bibfield  {journal} {\bibinfo
  {journal} {Classical and Quantum Gravity}\ }\textbf {\bibinfo {volume}
  {36}},\ \bibinfo {pages} {034001} (\bibinfo {year} {2019})}\BibitemShut
  {NoStop}%
\bibitem [{\citenamefont {Bose}\ \emph {et~al.}(2023)\citenamefont {Bose},
  \citenamefont {Fuentes}, \citenamefont {Geraci}, \citenamefont {Khan},
  \citenamefont {Qvarfort}, \citenamefont {Rademacher}, \citenamefont {Rashid},
  \citenamefont {Toro{\v{s}}}, \citenamefont {Ulbricht},\ and\ \citenamefont
  {Wanjura}}]{bose2023massive}%
  \BibitemOpen
  \bibfield  {author} {\bibinfo {author} {\bibfnamefont {S.}~\bibnamefont
  {Bose}}, \bibinfo {author} {\bibfnamefont {I.}~\bibnamefont {Fuentes}},
  \bibinfo {author} {\bibfnamefont {A.~A.}\ \bibnamefont {Geraci}}, \bibinfo
  {author} {\bibfnamefont {S.~M.}\ \bibnamefont {Khan}}, \bibinfo {author}
  {\bibfnamefont {S.}~\bibnamefont {Qvarfort}}, \bibinfo {author}
  {\bibfnamefont {M.}~\bibnamefont {Rademacher}}, \bibinfo {author}
  {\bibfnamefont {M.}~\bibnamefont {Rashid}}, \bibinfo {author} {\bibfnamefont
  {M.}~\bibnamefont {Toro{\v{s}}}}, \bibinfo {author} {\bibfnamefont
  {H.}~\bibnamefont {Ulbricht}},\ and\ \bibinfo {author} {\bibfnamefont
  {C.~C.}\ \bibnamefont {Wanjura}},\ }\bibfield  {title} {\bibinfo {title}
  {Massive quantum systems as interfaces of quantum mechanics and gravity},\
  }\href {http://arxiv.org/abs/2311.09218} {\bibfield  {journal} {\bibinfo
  {journal} {arXiv Prepr.}\ }\textbf {\bibinfo {volume} {2311.09218}} (\bibinfo
  {year} {2023})}\BibitemShut {NoStop}%
\bibitem [{\citenamefont {Marletto}\ and\ \citenamefont
  {Vedral}(2024)}]{marletto2024quantum}%
  \BibitemOpen
  \bibfield  {author} {\bibinfo {author} {\bibfnamefont {C.}~\bibnamefont
  {Marletto}}\ and\ \bibinfo {author} {\bibfnamefont {V.}~\bibnamefont
  {Vedral}},\ }\bibfield  {title} {\bibinfo {title} {Quantum-information
  methods for quantum gravity laboratory-based tests},\ }\bibfield  {journal}
  {\bibinfo  {journal} {arXiv preprint arXiv:2410.07262}\ }\href
  {https://doi.org/10.48550/arXiv.2410.07262} {10.48550/arXiv.2410.07262}
  (\bibinfo {year} {2024})\BibitemShut {NoStop}%
\bibitem [{\citenamefont {Lami}\ \emph {et~al.}(2024)\citenamefont {Lami},
  \citenamefont {Pedernales},\ and\ \citenamefont {Plenio}}]{lami2024testing}%
  \BibitemOpen
  \bibfield  {author} {\bibinfo {author} {\bibfnamefont {L.}~\bibnamefont
  {Lami}}, \bibinfo {author} {\bibfnamefont {J.~S.}\ \bibnamefont
  {Pedernales}},\ and\ \bibinfo {author} {\bibfnamefont {M.~B.}\ \bibnamefont
  {Plenio}},\ }\bibfield  {title} {\bibinfo {title} {Testing the quantumness of
  gravity without entanglement},\ }\href
  {https://doi.org/10.1103/PhysRevX.14.021022} {\bibfield  {journal} {\bibinfo
  {journal} {Physical Review X}\ }\textbf {\bibinfo {volume} {14}},\ \bibinfo
  {pages} {021022} (\bibinfo {year} {2024})}\BibitemShut {NoStop}%
\bibitem [{\citenamefont {Tobar}\ \emph {et~al.}(2024)\citenamefont {Tobar},
  \citenamefont {Manikandan}, \citenamefont {Beitel},\ and\ \citenamefont
  {Pikovski}}]{tobar2024detecting}%
  \BibitemOpen
  \bibfield  {author} {\bibinfo {author} {\bibfnamefont {G.}~\bibnamefont
  {Tobar}}, \bibinfo {author} {\bibfnamefont {S.~K.}\ \bibnamefont
  {Manikandan}}, \bibinfo {author} {\bibfnamefont {T.}~\bibnamefont {Beitel}},\
  and\ \bibinfo {author} {\bibfnamefont {I.}~\bibnamefont {Pikovski}},\
  }\bibfield  {title} {\bibinfo {title} {Detecting single gravitons with
  quantum sensing},\ }\href
  {https://doi.org/https://doi.org/10.1038/s41467-024-51420-8} {\bibfield
  {journal} {\bibinfo  {journal} {Nature Communications}\ }\textbf {\bibinfo
  {volume} {15}},\ \bibinfo {pages} {7229} (\bibinfo {year}
  {2024})}\BibitemShut {NoStop}%
\bibitem [{\citenamefont {Colella}\ \emph {et~al.}(1975)\citenamefont
  {Colella}, \citenamefont {Overhauser},\ and\ \citenamefont
  {Werner}}]{colella1975observation}%
  \BibitemOpen
  \bibfield  {author} {\bibinfo {author} {\bibfnamefont {R.}~\bibnamefont
  {Colella}}, \bibinfo {author} {\bibfnamefont {A.~W.}\ \bibnamefont
  {Overhauser}},\ and\ \bibinfo {author} {\bibfnamefont {S.~A.}\ \bibnamefont
  {Werner}},\ }\bibfield  {title} {\bibinfo {title} {Observation of
  gravitationally induced quantum interference},\ }\href
  {https://doi.org/https://doi.org/10.1103/PhysRevLett.34.1472} {\bibfield
  {journal} {\bibinfo  {journal} {Physical Review Letters}\ }\textbf {\bibinfo
  {volume} {34}},\ \bibinfo {pages} {1472} (\bibinfo {year}
  {1975})}\BibitemShut {NoStop}%
\bibitem [{\citenamefont {Overstreet}\ \emph {et~al.}(2022)\citenamefont
  {Overstreet}, \citenamefont {Asenbaum}, \citenamefont {Curti}, \citenamefont
  {Kim},\ and\ \citenamefont {Kasevich}}]{overstreet2022observation}%
  \BibitemOpen
  \bibfield  {author} {\bibinfo {author} {\bibfnamefont {C.}~\bibnamefont
  {Overstreet}}, \bibinfo {author} {\bibfnamefont {P.}~\bibnamefont
  {Asenbaum}}, \bibinfo {author} {\bibfnamefont {J.}~\bibnamefont {Curti}},
  \bibinfo {author} {\bibfnamefont {M.}~\bibnamefont {Kim}},\ and\ \bibinfo
  {author} {\bibfnamefont {M.~A.}\ \bibnamefont {Kasevich}},\ }\bibfield
  {title} {\bibinfo {title} {Observation of a gravitational aharonov-bohm
  effect},\ }\href {https://doi.org/https://doi.org/10.1126/science.abl7152}
  {\bibfield  {journal} {\bibinfo  {journal} {Science}\ }\textbf {\bibinfo
  {volume} {375}},\ \bibinfo {pages} {226} (\bibinfo {year}
  {2022})}\BibitemShut {NoStop}%
\bibitem [{\citenamefont {Nesvizhevsky}\ \emph {et~al.}(2002)\citenamefont
  {Nesvizhevsky}, \citenamefont {B{\"o}rner}, \citenamefont {Petukhov},
  \citenamefont {Abele}, \citenamefont {Bae{\ss}ler}, \citenamefont {Rue{\ss}},
  \citenamefont {St{\"o}ferle}, \citenamefont {Westphal}, \citenamefont
  {Gagarski}, \citenamefont {Petrov} \emph {et~al.}}]{nesvizhevsky2002quantum}%
  \BibitemOpen
  \bibfield  {author} {\bibinfo {author} {\bibfnamefont {V.~V.}\ \bibnamefont
  {Nesvizhevsky}}, \bibinfo {author} {\bibfnamefont {H.~G.}\ \bibnamefont
  {B{\"o}rner}}, \bibinfo {author} {\bibfnamefont {A.~K.}\ \bibnamefont
  {Petukhov}}, \bibinfo {author} {\bibfnamefont {H.}~\bibnamefont {Abele}},
  \bibinfo {author} {\bibfnamefont {S.}~\bibnamefont {Bae{\ss}ler}}, \bibinfo
  {author} {\bibfnamefont {F.~J.}\ \bibnamefont {Rue{\ss}}}, \bibinfo {author}
  {\bibfnamefont {T.}~\bibnamefont {St{\"o}ferle}}, \bibinfo {author}
  {\bibfnamefont {A.}~\bibnamefont {Westphal}}, \bibinfo {author}
  {\bibfnamefont {A.~M.}\ \bibnamefont {Gagarski}}, \bibinfo {author}
  {\bibfnamefont {G.~A.}\ \bibnamefont {Petrov}}, \emph {et~al.},\ }\bibfield
  {title} {\bibinfo {title} {Quantum states of neutrons in the earth's
  gravitational field},\ }\href
  {https://doi.org/https://doi.org/10.1038/415297a} {\bibfield  {journal}
  {\bibinfo  {journal} {Nature}\ }\textbf {\bibinfo {volume} {415}},\ \bibinfo
  {pages} {297} (\bibinfo {year} {2002})}\BibitemShut {NoStop}%
\bibitem [{\citenamefont {Jenke}\ \emph {et~al.}(2009)\citenamefont {Jenke},
  \citenamefont {Stadler}, \citenamefont {Abele},\ and\ \citenamefont
  {Geltenbort}}]{jenke2009q}%
  \BibitemOpen
  \bibfield  {author} {\bibinfo {author} {\bibfnamefont {T.}~\bibnamefont
  {Jenke}}, \bibinfo {author} {\bibfnamefont {D.}~\bibnamefont {Stadler}},
  \bibinfo {author} {\bibfnamefont {H.}~\bibnamefont {Abele}},\ and\ \bibinfo
  {author} {\bibfnamefont {P.}~\bibnamefont {Geltenbort}},\ }\bibfield  {title}
  {\bibinfo {title} {Q-bounce—experiments with quantum bouncing ultracold
  neutrons},\ }\href
  {https://doi.org/https://doi.org/10.1016/j.nima.2009.07.073} {\bibfield
  {journal} {\bibinfo  {journal} {Nuclear Instruments and Methods in Physics
  Research Section A: Accelerators, Spectrometers, Detectors and Associated
  Equipment}\ }\textbf {\bibinfo {volume} {611}},\ \bibinfo {pages} {318}
  (\bibinfo {year} {2009})}\BibitemShut {NoStop}%
\bibitem [{\citenamefont {Dimopoulos}\ \emph {et~al.}(2008)\citenamefont
  {Dimopoulos}, \citenamefont {Graham}, \citenamefont {Hogan},\ and\
  \citenamefont {Kasevich}}]{dimopoulos2008general}%
  \BibitemOpen
  \bibfield  {author} {\bibinfo {author} {\bibfnamefont {S.}~\bibnamefont
  {Dimopoulos}}, \bibinfo {author} {\bibfnamefont {P.~W.}\ \bibnamefont
  {Graham}}, \bibinfo {author} {\bibfnamefont {J.~M.}\ \bibnamefont {Hogan}},\
  and\ \bibinfo {author} {\bibfnamefont {M.~A.}\ \bibnamefont {Kasevich}},\
  }\bibfield  {title} {\bibinfo {title} {General relativistic effects in atom
  interferometry},\ }\href
  {https://doi.org/https://doi.org/10.1103/PhysRevD.78.042003} {\bibfield
  {journal} {\bibinfo  {journal} {Physical Review D—Particles, Fields,
  Gravitation, and Cosmology}\ }\textbf {\bibinfo {volume} {78}},\ \bibinfo
  {pages} {042003} (\bibinfo {year} {2008})}\BibitemShut {NoStop}%
\bibitem [{\citenamefont {Zych}\ \emph {et~al.}(2011)\citenamefont {Zych},
  \citenamefont {Costa}, \citenamefont {Pikovski},\ and\ \citenamefont
  {Brukner}}]{Zych2011}%
  \BibitemOpen
  \bibfield  {author} {\bibinfo {author} {\bibfnamefont {M.}~\bibnamefont
  {Zych}}, \bibinfo {author} {\bibfnamefont {F.}~\bibnamefont {Costa}},
  \bibinfo {author} {\bibfnamefont {I.}~\bibnamefont {Pikovski}},\ and\
  \bibinfo {author} {\bibfnamefont {{\v C}.}~\bibnamefont {Brukner}},\
  }\bibfield  {title} {\bibinfo {title} {Quantum interferometric visibility as
  a witness of general relativistic proper time},\ }\href
  {http://dx.doi.org/10.1038/ncomms1498} {\bibfield  {journal} {\bibinfo
  {journal} {Nature Communications}\ }\textbf {\bibinfo {volume} {2}},\
  \bibinfo {pages} {505 EP } (\bibinfo {year} {2011})}\BibitemShut {NoStop}%
\bibitem [{\citenamefont {Sinha}\ and\ \citenamefont
  {Samuel}(2011)}]{sinha2011atom}%
  \BibitemOpen
  \bibfield  {author} {\bibinfo {author} {\bibfnamefont {S.}~\bibnamefont
  {Sinha}}\ and\ \bibinfo {author} {\bibfnamefont {J.}~\bibnamefont {Samuel}},\
  }\bibfield  {title} {\bibinfo {title} {Atom interferometry and the
  gravitational redshift},\ }\href
  {https://doi.org/10.1088/0264-9381/28/14/145018} {\bibfield  {journal}
  {\bibinfo  {journal} {Classical and Quantum Gravity}\ }\textbf {\bibinfo
  {volume} {28}},\ \bibinfo {pages} {145018} (\bibinfo {year}
  {2011})}\BibitemShut {NoStop}%
\bibitem [{\citenamefont {Pikovski}\ \emph {et~al.}(2015)\citenamefont
  {Pikovski}, \citenamefont {Zych}, \citenamefont {Costa},\ and\ \citenamefont
  {Brukner}}]{Pikovski2015}%
  \BibitemOpen
  \bibfield  {author} {\bibinfo {author} {\bibfnamefont {I.}~\bibnamefont
  {Pikovski}}, \bibinfo {author} {\bibfnamefont {M.}~\bibnamefont {Zych}},
  \bibinfo {author} {\bibfnamefont {F.}~\bibnamefont {Costa}},\ and\ \bibinfo
  {author} {\bibfnamefont {{\v{C}}.}~\bibnamefont {Brukner}},\ }\bibfield
  {title} {\bibinfo {title} {{Universal decoherence due to gravitational time
  dilation}},\ }\href {https://doi.org/10.1038/nphys3366} {\bibfield  {journal}
  {\bibinfo  {journal} {Nat. Phys.}\ }\textbf {\bibinfo {volume} {11}},\
  \bibinfo {pages} {668} (\bibinfo {year} {2015})}\BibitemShut {NoStop}%
\bibitem [{\citenamefont {Zych}\ \emph {et~al.}(2016)\citenamefont {Zych},
  \citenamefont {Pikovski}, \citenamefont {Costa},\ and\ \citenamefont
  {Brukner}}]{zych2016general}%
  \BibitemOpen
  \bibfield  {author} {\bibinfo {author} {\bibfnamefont {M.}~\bibnamefont
  {Zych}}, \bibinfo {author} {\bibfnamefont {I.}~\bibnamefont {Pikovski}},
  \bibinfo {author} {\bibfnamefont {F.}~\bibnamefont {Costa}},\ and\ \bibinfo
  {author} {\bibfnamefont {{\v{C}}.}~\bibnamefont {Brukner}},\ }\bibfield
  {title} {\bibinfo {title} {General relativistic effects in quantum
  interference of “clocks”},\ }in\ \href
  {https://doi.org/10.1088/1742-6596/723/1/012044} {\emph {\bibinfo {booktitle}
  {Journal of Physics: Conference Series}}},\ Vol.\ \bibinfo {volume} {723}\
  (\bibinfo {organization} {IOP Publishing},\ \bibinfo {year} {2016})\ p.\
  \bibinfo {pages} {012044}\BibitemShut {NoStop}%
\bibitem [{\citenamefont {Orlando}\ \emph {et~al.}(2017)\citenamefont
  {Orlando}, \citenamefont {Pollock},\ and\ \citenamefont
  {Modi}}]{orlando2017does}%
  \BibitemOpen
  \bibfield  {author} {\bibinfo {author} {\bibfnamefont {P.~J.}\ \bibnamefont
  {Orlando}}, \bibinfo {author} {\bibfnamefont {F.~A.}\ \bibnamefont
  {Pollock}},\ and\ \bibinfo {author} {\bibfnamefont {K.}~\bibnamefont
  {Modi}},\ }\bibfield  {title} {\bibinfo {title} {How does interference
  fall?},\ }\href {https://doi.org/10.1007/978-3-319-53412-1_19} {\bibfield
  {journal} {\bibinfo  {journal} {Lectures on general quantum correlations and
  their applications}\ ,\ \bibinfo {pages} {421}} (\bibinfo {year}
  {2017})}\BibitemShut {NoStop}%
\bibitem [{\citenamefont {Zhou}\ \emph {et~al.}(2018)\citenamefont {Zhou},
  \citenamefont {Margalit}, \citenamefont {Rohrlich}, \citenamefont {Japha},\
  and\ \citenamefont {Folman}}]{zhou2018quantum}%
  \BibitemOpen
  \bibfield  {author} {\bibinfo {author} {\bibfnamefont {Z.}~\bibnamefont
  {Zhou}}, \bibinfo {author} {\bibfnamefont {Y.}~\bibnamefont {Margalit}},
  \bibinfo {author} {\bibfnamefont {D.}~\bibnamefont {Rohrlich}}, \bibinfo
  {author} {\bibfnamefont {Y.}~\bibnamefont {Japha}},\ and\ \bibinfo {author}
  {\bibfnamefont {R.}~\bibnamefont {Folman}},\ }\bibfield  {title} {\bibinfo
  {title} {Quantum complementarity of clocks in the context of general
  relativity},\ }\href {https://doi.org/10.1088/1361-6382/aad56b} {\bibfield
  {journal} {\bibinfo  {journal} {Classical and quantum gravity}\ }\textbf
  {\bibinfo {volume} {35}},\ \bibinfo {pages} {185003} (\bibinfo {year}
  {2018})}\BibitemShut {NoStop}%
\bibitem [{\citenamefont {Loriani}\ \emph {et~al.}(2019)\citenamefont
  {Loriani}, \citenamefont {Friedrich}, \citenamefont {Ufrecht}, \citenamefont
  {Di~Pumpo}, \citenamefont {Kleinert}, \citenamefont {Abend}, \citenamefont
  {Gaaloul}, \citenamefont {Meiners}, \citenamefont {Schubert}, \citenamefont
  {Tell} \emph {et~al.}}]{loriani2019interference}%
  \BibitemOpen
  \bibfield  {author} {\bibinfo {author} {\bibfnamefont {S.}~\bibnamefont
  {Loriani}}, \bibinfo {author} {\bibfnamefont {A.}~\bibnamefont {Friedrich}},
  \bibinfo {author} {\bibfnamefont {C.}~\bibnamefont {Ufrecht}}, \bibinfo
  {author} {\bibfnamefont {F.}~\bibnamefont {Di~Pumpo}}, \bibinfo {author}
  {\bibfnamefont {S.}~\bibnamefont {Kleinert}}, \bibinfo {author}
  {\bibfnamefont {S.}~\bibnamefont {Abend}}, \bibinfo {author} {\bibfnamefont
  {N.}~\bibnamefont {Gaaloul}}, \bibinfo {author} {\bibfnamefont
  {C.}~\bibnamefont {Meiners}}, \bibinfo {author} {\bibfnamefont
  {C.}~\bibnamefont {Schubert}}, \bibinfo {author} {\bibfnamefont
  {D.}~\bibnamefont {Tell}}, \emph {et~al.},\ }\bibfield  {title} {\bibinfo
  {title} {Interference of clocks: A quantum twin paradox},\ }\href
  {https://doi.org/DOI: 10.1126/sciadv.aax8966} {\bibfield  {journal} {\bibinfo
   {journal} {Science advances}\ }\textbf {\bibinfo {volume} {5}},\ \bibinfo
  {pages} {eaax8966} (\bibinfo {year} {2019})}\BibitemShut {NoStop}%
\bibitem [{\citenamefont {Roura}(2020)}]{roura2020gravitational}%
  \BibitemOpen
  \bibfield  {author} {\bibinfo {author} {\bibfnamefont {A.}~\bibnamefont
  {Roura}},\ }\bibfield  {title} {\bibinfo {title} {Gravitational redshift in
  quantum-clock interferometry},\ }\href
  {https://doi.org/10.1103/PhysRevX.10.021014} {\bibfield  {journal} {\bibinfo
  {journal} {Physical review X}\ }\textbf {\bibinfo {volume} {10}},\ \bibinfo
  {pages} {021014} (\bibinfo {year} {2020})}\BibitemShut {NoStop}%
\bibitem [{\citenamefont {Margalit}\ \emph {et~al.}(2015)\citenamefont
  {Margalit}, \citenamefont {Zhou}, \citenamefont {Machluf}, \citenamefont
  {Rohrlich}, \citenamefont {Japha},\ and\ \citenamefont
  {Folman}}]{margalit2015self}%
  \BibitemOpen
  \bibfield  {author} {\bibinfo {author} {\bibfnamefont {Y.}~\bibnamefont
  {Margalit}}, \bibinfo {author} {\bibfnamefont {Z.}~\bibnamefont {Zhou}},
  \bibinfo {author} {\bibfnamefont {S.}~\bibnamefont {Machluf}}, \bibinfo
  {author} {\bibfnamefont {D.}~\bibnamefont {Rohrlich}}, \bibinfo {author}
  {\bibfnamefont {Y.}~\bibnamefont {Japha}},\ and\ \bibinfo {author}
  {\bibfnamefont {R.}~\bibnamefont {Folman}},\ }\bibfield  {title} {\bibinfo
  {title} {A self-interfering clock as a “which path” witness},\ }\href
  {https://doi.org/10.1126/science.aac6498} {\bibfield  {journal} {\bibinfo
  {journal} {Science}\ }\textbf {\bibinfo {volume} {349}},\ \bibinfo {pages}
  {1205} (\bibinfo {year} {2015})}\BibitemShut {NoStop}%
\bibitem [{\citenamefont {Barzel}\ \emph {et~al.}(2024)\citenamefont {Barzel},
  \citenamefont {G{\"u}ndo{\u{g}}an}, \citenamefont {Krutzik}, \citenamefont
  {R{\"a}tzel},\ and\ \citenamefont {L{\"a}mmerzahl}}]{barzel2024entanglement}%
  \BibitemOpen
  \bibfield  {author} {\bibinfo {author} {\bibfnamefont {R.}~\bibnamefont
  {Barzel}}, \bibinfo {author} {\bibfnamefont {M.}~\bibnamefont
  {G{\"u}ndo{\u{g}}an}}, \bibinfo {author} {\bibfnamefont {M.}~\bibnamefont
  {Krutzik}}, \bibinfo {author} {\bibfnamefont {D.}~\bibnamefont
  {R{\"a}tzel}},\ and\ \bibinfo {author} {\bibfnamefont {C.}~\bibnamefont
  {L{\"a}mmerzahl}},\ }\bibfield  {title} {\bibinfo {title} {Entanglement
  dynamics of photon pairs and quantum memories in the gravitational field of
  the earth},\ }\href {https://doi.org/10.22331/q-2024-02-29-1273} {\bibfield
  {journal} {\bibinfo  {journal} {Quantum}\ }\textbf {\bibinfo {volume} {8}},\
  \bibinfo {pages} {1273} (\bibinfo {year} {2024})}\BibitemShut {NoStop}%
\bibitem [{\citenamefont {Borregaard}\ and\ \citenamefont
  {Pikovski}(2024)}]{borregaard2024testing}%
  \BibitemOpen
  \bibfield  {author} {\bibinfo {author} {\bibfnamefont {J.}~\bibnamefont
  {Borregaard}}\ and\ \bibinfo {author} {\bibfnamefont {I.}~\bibnamefont
  {Pikovski}},\ }\bibfield  {title} {\bibinfo {title} {Testing quantum theory
  on curved space-time with quantum networks},\ }\href
  {http://arxiv.org/abs/2406.19533} {\bibfield  {journal} {\bibinfo  {journal}
  {arXiv Prepr.}\ }\textbf {\bibinfo {volume} {2406.19533}} (\bibinfo {year}
  {2024})}\BibitemShut {NoStop}%
\bibitem [{\citenamefont {Covey}\ \emph {et~al.}(2019)\citenamefont {Covey},
  \citenamefont {Sipahigil}, \citenamefont {Szoke}, \citenamefont {Sinclair},
  \citenamefont {Endres},\ and\ \citenamefont {Painter}}]{Covey2019b}%
  \BibitemOpen
  \bibfield  {author} {\bibinfo {author} {\bibfnamefont {J.~P.}\ \bibnamefont
  {Covey}}, \bibinfo {author} {\bibfnamefont {A.}~\bibnamefont {Sipahigil}},
  \bibinfo {author} {\bibfnamefont {S.}~\bibnamefont {Szoke}}, \bibinfo
  {author} {\bibfnamefont {N.}~\bibnamefont {Sinclair}}, \bibinfo {author}
  {\bibfnamefont {M.}~\bibnamefont {Endres}},\ and\ \bibinfo {author}
  {\bibfnamefont {O.}~\bibnamefont {Painter}},\ }\bibfield  {title} {\bibinfo
  {title} {{Telecom-Band Quantum Optics with Ytterbium Atoms and Silicon
  Nanophotonics}},\ }\href {https://doi.org/10.1103/PhysRevApplied.11.034044}
  {\bibfield  {journal} {\bibinfo  {journal} {Phys. Rev. Appl.}\ }\textbf
  {\bibinfo {volume} {11}},\ \bibinfo {pages} {034044} (\bibinfo {year}
  {2019})}\BibitemShut {NoStop}%
\bibitem [{\citenamefont {Huie}\ \emph {et~al.}(2021)\citenamefont {Huie},
  \citenamefont {Menon}, \citenamefont {Bernien},\ and\ \citenamefont
  {Covey}}]{Huie2021}%
  \BibitemOpen
  \bibfield  {author} {\bibinfo {author} {\bibfnamefont {W.}~\bibnamefont
  {Huie}}, \bibinfo {author} {\bibfnamefont {S.~G.}\ \bibnamefont {Menon}},
  \bibinfo {author} {\bibfnamefont {H.}~\bibnamefont {Bernien}},\ and\ \bibinfo
  {author} {\bibfnamefont {J.~P.}\ \bibnamefont {Covey}},\ }\bibfield  {title}
  {\bibinfo {title} {{Multiplexed telecommunication-band quantum networking
  with atom arrays in optical cavities}},\ }\href
  {https://doi.org/10.1103/PhysRevResearch.3.043154} {\bibfield  {journal}
  {\bibinfo  {journal} {Phys. Rev. Res.}\ }\textbf {\bibinfo {volume} {3}},\
  \bibinfo {pages} {043154} (\bibinfo {year} {2021})}\BibitemShut {NoStop}%
\bibitem [{\citenamefont {Li}\ and\ \citenamefont
  {Thompson}(2024)}]{LiThompson2024}%
  \BibitemOpen
  \bibfield  {author} {\bibinfo {author} {\bibfnamefont {Y.}~\bibnamefont
  {Li}}\ and\ \bibinfo {author} {\bibfnamefont {J.~D.}\ \bibnamefont
  {Thompson}},\ }\bibfield  {title} {\bibinfo {title} {{High-Rate and
  High-Fidelity Modular Interconnects between Neutral Atom Quantum
  Processors}},\ }\href {https://doi.org/10.1103/PRXQuantum.5.020363}
  {\bibfield  {journal} {\bibinfo  {journal} {PRX Quantum}\ }\textbf {\bibinfo
  {volume} {5}},\ \bibinfo {pages} {020363} (\bibinfo {year}
  {2024})}\BibitemShut {NoStop}%
\bibitem [{\citenamefont {Sunami}\ \emph {et~al.}(2024)\citenamefont {Sunami},
  \citenamefont {Tamiya}, \citenamefont {Inoue}, \citenamefont {Yamasaki},\
  and\ \citenamefont {Goban}}]{Sunami2024}%
  \BibitemOpen
  \bibfield  {author} {\bibinfo {author} {\bibfnamefont {S.}~\bibnamefont
  {Sunami}}, \bibinfo {author} {\bibfnamefont {S.}~\bibnamefont {Tamiya}},
  \bibinfo {author} {\bibfnamefont {R.}~\bibnamefont {Inoue}}, \bibinfo
  {author} {\bibfnamefont {H.}~\bibnamefont {Yamasaki}},\ and\ \bibinfo
  {author} {\bibfnamefont {A.}~\bibnamefont {Goban}},\ }\bibfield  {title}
  {\bibinfo {title} {{Scalable Networking of Neutral-Atom Qubits:
  Nanofiber-Based Approach for Multiprocessor Fault-Tolerant Quantum
  Computer}},\ }\href {http://arxiv.org/abs/2407.11111} {\bibfield  {journal}
  {\bibinfo  {journal} {arXiv Prepr.}\ }\textbf {\bibinfo {volume}
  {2407.11111}} (\bibinfo {year} {2024})}\BibitemShut {NoStop}%
\bibitem [{\citenamefont {Hofmann}\ \emph {et~al.}(2012)\citenamefont
  {Hofmann}, \citenamefont {Krug}, \citenamefont {Ortegel}, \citenamefont
  {G{\'{e}}rard}, \citenamefont {Weber}, \citenamefont {Rosenfeld},\ and\
  \citenamefont {Weinfurter}}]{Hoffmann2012}%
  \BibitemOpen
  \bibfield  {author} {\bibinfo {author} {\bibfnamefont {J.}~\bibnamefont
  {Hofmann}}, \bibinfo {author} {\bibfnamefont {M.}~\bibnamefont {Krug}},
  \bibinfo {author} {\bibfnamefont {N.}~\bibnamefont {Ortegel}}, \bibinfo
  {author} {\bibfnamefont {L.}~\bibnamefont {G{\'{e}}rard}}, \bibinfo {author}
  {\bibfnamefont {M.}~\bibnamefont {Weber}}, \bibinfo {author} {\bibfnamefont
  {W.}~\bibnamefont {Rosenfeld}},\ and\ \bibinfo {author} {\bibfnamefont
  {H.}~\bibnamefont {Weinfurter}},\ }\bibfield  {title} {\bibinfo {title}
  {{Heralded Entanglement Between Widely Separated Atoms}},\ }\href
  {https://doi.org/10.1126/science.1221856} {\bibfield  {journal} {\bibinfo
  {journal} {Science}\ }\textbf {\bibinfo {volume} {337}},\ \bibinfo {pages}
  {72} (\bibinfo {year} {2012})}\BibitemShut {NoStop}%
\bibitem [{\citenamefont {Daiss}\ \emph {et~al.}(2021)\citenamefont {Daiss},
  \citenamefont {Langenfeld}, \citenamefont {Welte}, \citenamefont {Distante},
  \citenamefont {Thomas}, \citenamefont {Hartung}, \citenamefont {Morin},\ and\
  \citenamefont {Rempe}}]{Daiss2021}%
  \BibitemOpen
  \bibfield  {author} {\bibinfo {author} {\bibfnamefont {S.}~\bibnamefont
  {Daiss}}, \bibinfo {author} {\bibfnamefont {S.}~\bibnamefont {Langenfeld}},
  \bibinfo {author} {\bibfnamefont {S.}~\bibnamefont {Welte}}, \bibinfo
  {author} {\bibfnamefont {E.}~\bibnamefont {Distante}}, \bibinfo {author}
  {\bibfnamefont {P.}~\bibnamefont {Thomas}}, \bibinfo {author} {\bibfnamefont
  {L.}~\bibnamefont {Hartung}}, \bibinfo {author} {\bibfnamefont
  {O.}~\bibnamefont {Morin}},\ and\ \bibinfo {author} {\bibfnamefont
  {G.}~\bibnamefont {Rempe}},\ }\bibfield  {title} {\bibinfo {title} {{A
  quantum-logic gate between distant quantum-network modules}},\ }\href
  {https://doi.org/10.1126/science.abe3150} {\bibfield  {journal} {\bibinfo
  {journal} {Science}\ }\textbf {\bibinfo {volume} {371}},\ \bibinfo {pages}
  {614} (\bibinfo {year} {2021})}\BibitemShut {NoStop}%
\bibitem [{\citenamefont {Covey}\ \emph {et~al.}(2023)\citenamefont {Covey},
  \citenamefont {Weinfurter},\ and\ \citenamefont {Bernien}}]{Covey2023}%
  \BibitemOpen
  \bibfield  {author} {\bibinfo {author} {\bibfnamefont {J.~P.}\ \bibnamefont
  {Covey}}, \bibinfo {author} {\bibfnamefont {H.}~\bibnamefont {Weinfurter}},\
  and\ \bibinfo {author} {\bibfnamefont {H.}~\bibnamefont {Bernien}},\
  }\bibfield  {title} {\bibinfo {title} {{Quantum networks with neutral atom
  processing nodes}},\ }\href {https://doi.org/10.1038/s41534-023-00759-9}
  {\bibfield  {journal} {\bibinfo  {journal} {npj Quantum Inf.}\ }\textbf
  {\bibinfo {volume} {9}},\ \bibinfo {pages} {90} (\bibinfo {year}
  {2023})}\BibitemShut {NoStop}%
\bibitem [{\citenamefont {Hartung}\ \emph {et~al.}(2024)\citenamefont
  {Hartung}, \citenamefont {Seubert}, \citenamefont {Welte}, \citenamefont
  {Distante},\ and\ \citenamefont {Rempe}}]{Hartung2024}%
  \BibitemOpen
  \bibfield  {author} {\bibinfo {author} {\bibfnamefont {L.}~\bibnamefont
  {Hartung}}, \bibinfo {author} {\bibfnamefont {M.}~\bibnamefont {Seubert}},
  \bibinfo {author} {\bibfnamefont {S.}~\bibnamefont {Welte}}, \bibinfo
  {author} {\bibfnamefont {E.}~\bibnamefont {Distante}},\ and\ \bibinfo
  {author} {\bibfnamefont {G.}~\bibnamefont {Rempe}},\ }\bibfield  {title}
  {\bibinfo {title} {{A quantum-network register assembled with optical
  tweezers in an optical cavity}},\ }\href
  {https://doi.org/10.1126/science.ado6471} {\bibfield  {journal} {\bibinfo
  {journal} {Science}\ }\textbf {\bibinfo {volume} {385}},\ \bibinfo {pages}
  {179} (\bibinfo {year} {2024})}\BibitemShut {NoStop}%
\bibitem [{\citenamefont {Li}\ \emph {et~al.}(2025)\citenamefont {Li},
  \citenamefont {Hu}, \citenamefont {Jia}, \citenamefont {Huie}, \citenamefont
  {Sun}, \citenamefont {Aakash}, \citenamefont {Dong}, \citenamefont
  {Hiri-O-Tuppa},\ and\ \citenamefont {Covey}}]{Li2025}%
  \BibitemOpen
  \bibfield  {author} {\bibinfo {author} {\bibfnamefont {L.}~\bibnamefont
  {Li}}, \bibinfo {author} {\bibfnamefont {S.}~\bibnamefont {Hu}}, \bibinfo
  {author} {\bibfnamefont {Z.}~\bibnamefont {Jia}}, \bibinfo {author}
  {\bibfnamefont {W.}~\bibnamefont {Huie}}, \bibinfo {author} {\bibfnamefont
  {W.~K.~C.}\ \bibnamefont {Sun}}, \bibinfo {author} {\bibnamefont {Aakash}},
  \bibinfo {author} {\bibfnamefont {M.}~\bibnamefont {Dong}}, \bibinfo {author}
  {\bibfnamefont {T.}~\bibnamefont {Hiri-O-Tuppa}},\ and\ \bibinfo {author}
  {\bibfnamefont {J.~P.}\ \bibnamefont {Covey}},\ }\bibfield  {title} {\bibinfo
  {title} {{Parallelized telecom quantum networking with a ytterbium-171 atom
  array}},\ }\href@noop {} {\bibfield  {journal} {\bibinfo  {journal} {To be
  posted}\ } (\bibinfo {year} {2025})}\BibitemShut {NoStop}%
\bibitem [{\citenamefont {Olmschenk}\ \emph {et~al.}(2009)\citenamefont
  {Olmschenk}, \citenamefont {Matsukevich}, \citenamefont {Maunz},
  \citenamefont {Hayes}, \citenamefont {Duan},\ and\ \citenamefont
  {Monroe}}]{Olmschenk2009}%
  \BibitemOpen
  \bibfield  {author} {\bibinfo {author} {\bibfnamefont {S.}~\bibnamefont
  {Olmschenk}}, \bibinfo {author} {\bibfnamefont {D.~N.}\ \bibnamefont
  {Matsukevich}}, \bibinfo {author} {\bibfnamefont {P.}~\bibnamefont {Maunz}},
  \bibinfo {author} {\bibfnamefont {D.}~\bibnamefont {Hayes}}, \bibinfo
  {author} {\bibfnamefont {L.-M.}\ \bibnamefont {Duan}},\ and\ \bibinfo
  {author} {\bibfnamefont {C.}~\bibnamefont {Monroe}},\ }\bibfield  {title}
  {\bibinfo {title} {{Quantum Teleportation Between Distant Matter Qubits}},\
  }\href {https://doi.org/10.1126/science.1167209} {\bibfield  {journal}
  {\bibinfo  {journal} {Science}\ }\textbf {\bibinfo {volume} {323}},\ \bibinfo
  {pages} {486} (\bibinfo {year} {2009})}\BibitemShut {NoStop}%
\bibitem [{\citenamefont {Stephenson}\ \emph {et~al.}(2020)\citenamefont
  {Stephenson}, \citenamefont {Nadlinger}, \citenamefont {Nichol},
  \citenamefont {An}, \citenamefont {Drmota}, \citenamefont {Ballance},
  \citenamefont {Thirumalai}, \citenamefont {Goodwin}, \citenamefont {Lucas},\
  and\ \citenamefont {Ballance}}]{Stephenson2020}%
  \BibitemOpen
  \bibfield  {author} {\bibinfo {author} {\bibfnamefont {L.~J.}\ \bibnamefont
  {Stephenson}}, \bibinfo {author} {\bibfnamefont {D.~P.}\ \bibnamefont
  {Nadlinger}}, \bibinfo {author} {\bibfnamefont {B.~C.}\ \bibnamefont
  {Nichol}}, \bibinfo {author} {\bibfnamefont {S.}~\bibnamefont {An}}, \bibinfo
  {author} {\bibfnamefont {P.}~\bibnamefont {Drmota}}, \bibinfo {author}
  {\bibfnamefont {T.~G.}\ \bibnamefont {Ballance}}, \bibinfo {author}
  {\bibfnamefont {K.}~\bibnamefont {Thirumalai}}, \bibinfo {author}
  {\bibfnamefont {J.~F.}\ \bibnamefont {Goodwin}}, \bibinfo {author}
  {\bibfnamefont {D.~M.}\ \bibnamefont {Lucas}},\ and\ \bibinfo {author}
  {\bibfnamefont {C.~J.}\ \bibnamefont {Ballance}},\ }\bibfield  {title}
  {\bibinfo {title} {{High-Rate, High-Fidelity Entanglement of Qubits Across an
  Elementary Quantum Network}},\ }\href
  {https://doi.org/10.1103/PhysRevLett.124.110501} {\bibfield  {journal}
  {\bibinfo  {journal} {Phys. Rev. Lett.}\ }\textbf {\bibinfo {volume} {124}},\
  \bibinfo {pages} {110501} (\bibinfo {year} {2020})}\BibitemShut {NoStop}%
\bibitem [{\citenamefont {Nichol}\ \emph {et~al.}(2022)\citenamefont {Nichol},
  \citenamefont {Srinivas}, \citenamefont {Nadlinger}, \citenamefont {Drmota},
  \citenamefont {Main}, \citenamefont {Araneda}, \citenamefont {Ballance},\
  and\ \citenamefont {Lucas}}]{Nichol2022}%
  \BibitemOpen
  \bibfield  {author} {\bibinfo {author} {\bibfnamefont {B.~C.}\ \bibnamefont
  {Nichol}}, \bibinfo {author} {\bibfnamefont {R.}~\bibnamefont {Srinivas}},
  \bibinfo {author} {\bibfnamefont {D.~P.}\ \bibnamefont {Nadlinger}}, \bibinfo
  {author} {\bibfnamefont {P.}~\bibnamefont {Drmota}}, \bibinfo {author}
  {\bibfnamefont {D.}~\bibnamefont {Main}}, \bibinfo {author} {\bibfnamefont
  {G.}~\bibnamefont {Araneda}}, \bibinfo {author} {\bibfnamefont {C.~J.}\
  \bibnamefont {Ballance}},\ and\ \bibinfo {author} {\bibfnamefont {D.~M.}\
  \bibnamefont {Lucas}},\ }\bibfield  {title} {\bibinfo {title} {{An elementary
  quantum network of entangled optical atomic clocks}},\ }\href
  {https://doi.org/10.1038/s41586-022-05088-z} {\bibfield  {journal} {\bibinfo
  {journal} {Nature}\ }\textbf {\bibinfo {volume} {609}},\ \bibinfo {pages}
  {689} (\bibinfo {year} {2022})}\BibitemShut {NoStop}%
\bibitem [{\citenamefont {Saha}\ \emph {et~al.}(2024)\citenamefont {Saha},
  \citenamefont {Shalaev}, \citenamefont {O'Reilly}, \citenamefont {Goetting},
  \citenamefont {Toh}, \citenamefont {Kalakuntla}, \citenamefont {Yu},\ and\
  \citenamefont {Monroe}}]{Saha2024}%
  \BibitemOpen
  \bibfield  {author} {\bibinfo {author} {\bibfnamefont {S.}~\bibnamefont
  {Saha}}, \bibinfo {author} {\bibfnamefont {M.}~\bibnamefont {Shalaev}},
  \bibinfo {author} {\bibfnamefont {J.}~\bibnamefont {O'Reilly}}, \bibinfo
  {author} {\bibfnamefont {I.}~\bibnamefont {Goetting}}, \bibinfo {author}
  {\bibfnamefont {G.}~\bibnamefont {Toh}}, \bibinfo {author} {\bibfnamefont
  {A.}~\bibnamefont {Kalakuntla}}, \bibinfo {author} {\bibfnamefont
  {Y.}~\bibnamefont {Yu}},\ and\ \bibinfo {author} {\bibfnamefont
  {C.}~\bibnamefont {Monroe}},\ }\bibfield  {title} {\bibinfo {title}
  {{High-fidelity remote entanglement of trapped atoms mediated by time-bin
  photons}},\ }\href {http://arxiv.org/abs/2406.01761} {\bibfield  {journal}
  {\bibinfo  {journal} {arXiv Prepr.}\ }\textbf {\bibinfo {volume}
  {2406.01761}} (\bibinfo {year} {2024})},\ \Eprint
  {https://arxiv.org/abs/2406.01761} {arXiv:2406.01761} \BibitemShut {NoStop}%
\bibitem [{\citenamefont {Bernien}\ \emph {et~al.}(2013)\citenamefont
  {Bernien}, \citenamefont {Hensen}, \citenamefont {Pfaff}, \citenamefont
  {Koolstra}, \citenamefont {Blok}, \citenamefont {Robledo}, \citenamefont
  {Taminiau}, \citenamefont {Markham}, \citenamefont {Twitchen}, \citenamefont
  {Childress},\ and\ \citenamefont {Hanson}}]{Bernien2013}%
  \BibitemOpen
  \bibfield  {author} {\bibinfo {author} {\bibfnamefont {H.}~\bibnamefont
  {Bernien}}, \bibinfo {author} {\bibfnamefont {B.}~\bibnamefont {Hensen}},
  \bibinfo {author} {\bibfnamefont {W.}~\bibnamefont {Pfaff}}, \bibinfo
  {author} {\bibfnamefont {G.}~\bibnamefont {Koolstra}}, \bibinfo {author}
  {\bibfnamefont {M.~S.}\ \bibnamefont {Blok}}, \bibinfo {author}
  {\bibfnamefont {L.}~\bibnamefont {Robledo}}, \bibinfo {author} {\bibfnamefont
  {T.~H.}\ \bibnamefont {Taminiau}}, \bibinfo {author} {\bibfnamefont
  {M.}~\bibnamefont {Markham}}, \bibinfo {author} {\bibfnamefont {D.~J.}\
  \bibnamefont {Twitchen}}, \bibinfo {author} {\bibfnamefont {L.}~\bibnamefont
  {Childress}},\ and\ \bibinfo {author} {\bibfnamefont {R.}~\bibnamefont
  {Hanson}},\ }\bibfield  {title} {\bibinfo {title} {{Heralded entanglement
  between solid-state qubits separated by three metres}},\ }\href
  {https://doi.org/10.1038/nature12016} {\bibfield  {journal} {\bibinfo
  {journal} {Nature}\ }\textbf {\bibinfo {volume} {497}},\ \bibinfo {pages}
  {86} (\bibinfo {year} {2013})}\BibitemShut {NoStop}%
\bibitem [{\citenamefont {Humphreys}\ \emph {et~al.}(2018)\citenamefont
  {Humphreys}, \citenamefont {Kalb}, \citenamefont {Morits}, \citenamefont
  {Schouten}, \citenamefont {Vermeulen}, \citenamefont {Twitchen},
  \citenamefont {Markham},\ and\ \citenamefont {Hanson}}]{Humphreys2018}%
  \BibitemOpen
  \bibfield  {author} {\bibinfo {author} {\bibfnamefont {P.~C.}\ \bibnamefont
  {Humphreys}}, \bibinfo {author} {\bibfnamefont {N.}~\bibnamefont {Kalb}},
  \bibinfo {author} {\bibfnamefont {J.~P.~J.}\ \bibnamefont {Morits}}, \bibinfo
  {author} {\bibfnamefont {R.~N.}\ \bibnamefont {Schouten}}, \bibinfo {author}
  {\bibfnamefont {R.~F.~L.}\ \bibnamefont {Vermeulen}}, \bibinfo {author}
  {\bibfnamefont {D.~J.}\ \bibnamefont {Twitchen}}, \bibinfo {author}
  {\bibfnamefont {M.}~\bibnamefont {Markham}},\ and\ \bibinfo {author}
  {\bibfnamefont {R.}~\bibnamefont {Hanson}},\ }\bibfield  {title} {\bibinfo
  {title} {{Deterministic delivery of remote entanglement on a quantum
  network}},\ }\href {https://doi.org/10.1038/s41586-018-0200-5} {\bibfield
  {journal} {\bibinfo  {journal} {Nature}\ }\textbf {\bibinfo {volume} {558}},\
  \bibinfo {pages} {268} (\bibinfo {year} {2018})}\BibitemShut {NoStop}%
\bibitem [{\citenamefont {Kindem}\ \emph {et~al.}(2020)\citenamefont {Kindem},
  \citenamefont {Ruskuc}, \citenamefont {Bartholomew}, \citenamefont {Rochman},
  \citenamefont {Huan},\ and\ \citenamefont {Faraon}}]{Kindem2020}%
  \BibitemOpen
  \bibfield  {author} {\bibinfo {author} {\bibfnamefont {J.~M.}\ \bibnamefont
  {Kindem}}, \bibinfo {author} {\bibfnamefont {A.}~\bibnamefont {Ruskuc}},
  \bibinfo {author} {\bibfnamefont {J.~G.}\ \bibnamefont {Bartholomew}},
  \bibinfo {author} {\bibfnamefont {J.}~\bibnamefont {Rochman}}, \bibinfo
  {author} {\bibfnamefont {Y.~Q.}\ \bibnamefont {Huan}},\ and\ \bibinfo
  {author} {\bibfnamefont {A.}~\bibnamefont {Faraon}},\ }\bibfield  {title}
  {\bibinfo {title} {{Control and single-shot readout of an ion embedded in a
  nanophotonic cavity}},\ }\href {https://doi.org/10.1038/s41586-020-2160-9}
  {\bibfield  {journal} {\bibinfo  {journal} {Nature}\ }\textbf {\bibinfo
  {volume} {580}},\ \bibinfo {pages} {201} (\bibinfo {year}
  {2020})}\BibitemShut {NoStop}%
\bibitem [{\citenamefont {Bhaskar}\ \emph {et~al.}(2020)\citenamefont
  {Bhaskar}, \citenamefont {Riedinger}, \citenamefont {Machielse},
  \citenamefont {Levonian}, \citenamefont {Nguyen}, \citenamefont {Knall},
  \citenamefont {Park}, \citenamefont {Englund}, \citenamefont {Lon{\v{c}}ar},
  \citenamefont {Sukachev},\ and\ \citenamefont {Lukin}}]{Bhaskar2020}%
  \BibitemOpen
  \bibfield  {author} {\bibinfo {author} {\bibfnamefont {M.~K.}\ \bibnamefont
  {Bhaskar}}, \bibinfo {author} {\bibfnamefont {R.}~\bibnamefont {Riedinger}},
  \bibinfo {author} {\bibfnamefont {B.}~\bibnamefont {Machielse}}, \bibinfo
  {author} {\bibfnamefont {D.~S.}\ \bibnamefont {Levonian}}, \bibinfo {author}
  {\bibfnamefont {C.~T.}\ \bibnamefont {Nguyen}}, \bibinfo {author}
  {\bibfnamefont {E.~N.}\ \bibnamefont {Knall}}, \bibinfo {author}
  {\bibfnamefont {H.}~\bibnamefont {Park}}, \bibinfo {author} {\bibfnamefont
  {D.}~\bibnamefont {Englund}}, \bibinfo {author} {\bibfnamefont
  {M.}~\bibnamefont {Lon{\v{c}}ar}}, \bibinfo {author} {\bibfnamefont {D.~D.}\
  \bibnamefont {Sukachev}},\ and\ \bibinfo {author} {\bibfnamefont {M.~D.}\
  \bibnamefont {Lukin}},\ }\bibfield  {title} {\bibinfo {title} {{Experimental
  demonstration of memory-enhanced quantum communication}},\ }\href
  {https://doi.org/10.1038/s41586-020-2103-5} {\bibfield  {journal} {\bibinfo
  {journal} {Nature}\ }\textbf {\bibinfo {volume} {580}},\ \bibinfo {pages}
  {60} (\bibinfo {year} {2020})}\BibitemShut {NoStop}%
\bibitem [{\citenamefont {Uysal}\ \emph {et~al.}(2024)\citenamefont {Uysal},
  \citenamefont {Dusanowski}, \citenamefont {Xu}, \citenamefont {Horvath},
  \citenamefont {Ourari}, \citenamefont {Cava}, \citenamefont {de~Leon},\ and\
  \citenamefont {Thompson}}]{Uysal2024}%
  \BibitemOpen
  \bibfield  {author} {\bibinfo {author} {\bibfnamefont {M.~T.}\ \bibnamefont
  {Uysal}}, \bibinfo {author} {\bibfnamefont {{\L}.}~\bibnamefont
  {Dusanowski}}, \bibinfo {author} {\bibfnamefont {H.}~\bibnamefont {Xu}},
  \bibinfo {author} {\bibfnamefont {S.~P.}\ \bibnamefont {Horvath}}, \bibinfo
  {author} {\bibfnamefont {S.}~\bibnamefont {Ourari}}, \bibinfo {author}
  {\bibfnamefont {R.~J.}\ \bibnamefont {Cava}}, \bibinfo {author}
  {\bibfnamefont {N.~P.}\ \bibnamefont {de~Leon}},\ and\ \bibinfo {author}
  {\bibfnamefont {J.~D.}\ \bibnamefont {Thompson}},\ }\bibfield  {title}
  {\bibinfo {title} {{Spin-photon entanglement of a single Er$^{3+}$ ion in the
  telecom band}},\ }\href {http://arxiv.org/abs/2406.06515} {\bibfield
  {journal} {\bibinfo  {journal} {arXiv Prepr.}\ }\textbf {\bibinfo {volume}
  {2406.06515}} (\bibinfo {year} {2024})}\BibitemShut {NoStop}%
\bibitem [{\citenamefont {Lis}\ \emph {et~al.}(2023)\citenamefont {Lis},
  \citenamefont {Senoo}, \citenamefont {McGrew}, \citenamefont {R{\"{o}}nchen},
  \citenamefont {Jenkins},\ and\ \citenamefont {Kaufman}}]{Lis2023}%
  \BibitemOpen
  \bibfield  {author} {\bibinfo {author} {\bibfnamefont {J.~W.}\ \bibnamefont
  {Lis}}, \bibinfo {author} {\bibfnamefont {A.}~\bibnamefont {Senoo}}, \bibinfo
  {author} {\bibfnamefont {W.~F.}\ \bibnamefont {McGrew}}, \bibinfo {author}
  {\bibfnamefont {F.}~\bibnamefont {R{\"{o}}nchen}}, \bibinfo {author}
  {\bibfnamefont {A.}~\bibnamefont {Jenkins}},\ and\ \bibinfo {author}
  {\bibfnamefont {A.~M.}\ \bibnamefont {Kaufman}},\ }\bibfield  {title}
  {\bibinfo {title} {{Midcircuit Operations Using the omg Architecture in
  Neutral Atom Arrays}},\ }\href {https://doi.org/10.1103/PhysRevX.13.041035}
  {\bibfield  {journal} {\bibinfo  {journal} {Phys. Rev. X}\ }\textbf {\bibinfo
  {volume} {13}},\ \bibinfo {pages} {041035} (\bibinfo {year}
  {2023})}\BibitemShut {NoStop}%
\bibitem [{\citenamefont {Ma}\ \emph {et~al.}(2023)\citenamefont {Ma},
  \citenamefont {Liu}, \citenamefont {Peng}, \citenamefont {Zhang},
  \citenamefont {Jandura}, \citenamefont {Claes}, \citenamefont {Burgers},
  \citenamefont {Pupillo}, \citenamefont {Puri},\ and\ \citenamefont
  {Thompson}}]{Ma2023}%
  \BibitemOpen
  \bibfield  {author} {\bibinfo {author} {\bibfnamefont {S.}~\bibnamefont
  {Ma}}, \bibinfo {author} {\bibfnamefont {G.}~\bibnamefont {Liu}}, \bibinfo
  {author} {\bibfnamefont {P.}~\bibnamefont {Peng}}, \bibinfo {author}
  {\bibfnamefont {B.}~\bibnamefont {Zhang}}, \bibinfo {author} {\bibfnamefont
  {S.}~\bibnamefont {Jandura}}, \bibinfo {author} {\bibfnamefont
  {J.}~\bibnamefont {Claes}}, \bibinfo {author} {\bibfnamefont {A.~P.}\
  \bibnamefont {Burgers}}, \bibinfo {author} {\bibfnamefont {G.}~\bibnamefont
  {Pupillo}}, \bibinfo {author} {\bibfnamefont {S.}~\bibnamefont {Puri}},\ and\
  \bibinfo {author} {\bibfnamefont {J.~D.}\ \bibnamefont {Thompson}},\
  }\bibfield  {title} {\bibinfo {title} {{High-fidelity gates and mid-circuit
  erasure conversion in an atomic qubit}},\ }\href
  {https://doi.org/10.1038/s41586-023-06438-1} {\bibfield  {journal} {\bibinfo
  {journal} {Nature}\ }\textbf {\bibinfo {volume} {622}},\ \bibinfo {pages}
  {279} (\bibinfo {year} {2023})}\BibitemShut {NoStop}%
\bibitem [{\citenamefont {Muniz}\ \emph {et~al.}(2024)\citenamefont {Muniz},
  \citenamefont {Stone}, \citenamefont {Stack}, \citenamefont {Jaffe},
  \citenamefont {Kindem}, \citenamefont {Wadleigh}, \citenamefont
  {Zalys-Geller}, \citenamefont {Zhang}, \citenamefont {Chen}, \citenamefont
  {Norcia}, \citenamefont {Epstein}, \citenamefont {Halperin}, \citenamefont
  {Hummel}, \citenamefont {Wilkason}, \citenamefont {Li}, \citenamefont
  {Barnes}, \citenamefont {Battaglino}, \citenamefont {Bohdanowicz},
  \citenamefont {Booth}, \citenamefont {Brown}, \citenamefont {Brown},
  \citenamefont {Cairncross}, \citenamefont {Cassella}, \citenamefont {Coxe},
  \citenamefont {Crow}, \citenamefont {Feldkamp}, \citenamefont {Griger},
  \citenamefont {Heinz}, \citenamefont {Jones}, \citenamefont {Kim},
  \citenamefont {King}, \citenamefont {Kotru}, \citenamefont {Lauigan},
  \citenamefont {Marjanovic}, \citenamefont {Megidish}, \citenamefont
  {Meredith}, \citenamefont {McDonald}, \citenamefont {Morshead}, \citenamefont
  {Narayanaswami}, \citenamefont {Nishiguchi}, \citenamefont {Paule},
  \citenamefont {Pawlak}, \citenamefont {Pudenz}, \citenamefont {P{\'{e}}rez},
  \citenamefont {Ryou}, \citenamefont {Simon}, \citenamefont {Smull},
  \citenamefont {Urbanek}, \citenamefont {van~de Veerdonk}, \citenamefont
  {Vendeiro}, \citenamefont {Wu}, \citenamefont {Xie},\ and\ \citenamefont
  {Bloom}}]{Muniz2024}%
  \BibitemOpen
  \bibfield  {author} {\bibinfo {author} {\bibfnamefont {J.~A.}\ \bibnamefont
  {Muniz}}, \bibinfo {author} {\bibfnamefont {M.}~\bibnamefont {Stone}},
  \bibinfo {author} {\bibfnamefont {D.~T.}\ \bibnamefont {Stack}}, \bibinfo
  {author} {\bibfnamefont {M.}~\bibnamefont {Jaffe}}, \bibinfo {author}
  {\bibfnamefont {J.~M.}\ \bibnamefont {Kindem}}, \bibinfo {author}
  {\bibfnamefont {L.}~\bibnamefont {Wadleigh}}, \bibinfo {author}
  {\bibfnamefont {E.}~\bibnamefont {Zalys-Geller}}, \bibinfo {author}
  {\bibfnamefont {X.}~\bibnamefont {Zhang}}, \bibinfo {author} {\bibfnamefont
  {C.~A.}\ \bibnamefont {Chen}}, \bibinfo {author} {\bibfnamefont {M.~A.}\
  \bibnamefont {Norcia}}, \bibinfo {author} {\bibfnamefont {J.}~\bibnamefont
  {Epstein}}, \bibinfo {author} {\bibfnamefont {E.}~\bibnamefont {Halperin}},
  \bibinfo {author} {\bibfnamefont {F.}~\bibnamefont {Hummel}}, \bibinfo
  {author} {\bibfnamefont {T.}~\bibnamefont {Wilkason}}, \bibinfo {author}
  {\bibfnamefont {M.}~\bibnamefont {Li}}, \bibinfo {author} {\bibfnamefont
  {K.}~\bibnamefont {Barnes}}, \bibinfo {author} {\bibfnamefont
  {P.}~\bibnamefont {Battaglino}}, \bibinfo {author} {\bibfnamefont {T.~C.}\
  \bibnamefont {Bohdanowicz}}, \bibinfo {author} {\bibfnamefont
  {G.}~\bibnamefont {Booth}}, \bibinfo {author} {\bibfnamefont
  {A.}~\bibnamefont {Brown}}, \bibinfo {author} {\bibfnamefont {M.~O.}\
  \bibnamefont {Brown}}, \bibinfo {author} {\bibfnamefont {W.~B.}\ \bibnamefont
  {Cairncross}}, \bibinfo {author} {\bibfnamefont {K.}~\bibnamefont
  {Cassella}}, \bibinfo {author} {\bibfnamefont {R.}~\bibnamefont {Coxe}},
  \bibinfo {author} {\bibfnamefont {D.}~\bibnamefont {Crow}}, \bibinfo {author}
  {\bibfnamefont {M.}~\bibnamefont {Feldkamp}}, \bibinfo {author}
  {\bibfnamefont {C.}~\bibnamefont {Griger}}, \bibinfo {author} {\bibfnamefont
  {A.}~\bibnamefont {Heinz}}, \bibinfo {author} {\bibfnamefont {A.~M.~W.}\
  \bibnamefont {Jones}}, \bibinfo {author} {\bibfnamefont {H.}~\bibnamefont
  {Kim}}, \bibinfo {author} {\bibfnamefont {J.}~\bibnamefont {King}}, \bibinfo
  {author} {\bibfnamefont {K.}~\bibnamefont {Kotru}}, \bibinfo {author}
  {\bibfnamefont {J.}~\bibnamefont {Lauigan}}, \bibinfo {author} {\bibfnamefont
  {J.}~\bibnamefont {Marjanovic}}, \bibinfo {author} {\bibfnamefont
  {E.}~\bibnamefont {Megidish}}, \bibinfo {author} {\bibfnamefont
  {M.}~\bibnamefont {Meredith}}, \bibinfo {author} {\bibfnamefont
  {M.}~\bibnamefont {McDonald}}, \bibinfo {author} {\bibfnamefont
  {R.}~\bibnamefont {Morshead}}, \bibinfo {author} {\bibfnamefont
  {S.}~\bibnamefont {Narayanaswami}}, \bibinfo {author} {\bibfnamefont
  {C.}~\bibnamefont {Nishiguchi}}, \bibinfo {author} {\bibfnamefont
  {T.}~\bibnamefont {Paule}}, \bibinfo {author} {\bibfnamefont {K.~A.}\
  \bibnamefont {Pawlak}}, \bibinfo {author} {\bibfnamefont {K.~L.}\
  \bibnamefont {Pudenz}}, \bibinfo {author} {\bibfnamefont {D.~R.}\
  \bibnamefont {P{\'{e}}rez}}, \bibinfo {author} {\bibfnamefont
  {A.}~\bibnamefont {Ryou}}, \bibinfo {author} {\bibfnamefont {J.}~\bibnamefont
  {Simon}}, \bibinfo {author} {\bibfnamefont {A.}~\bibnamefont {Smull}},
  \bibinfo {author} {\bibfnamefont {M.}~\bibnamefont {Urbanek}}, \bibinfo
  {author} {\bibfnamefont {R.~J.~M.}\ \bibnamefont {van~de Veerdonk}}, \bibinfo
  {author} {\bibfnamefont {Z.}~\bibnamefont {Vendeiro}}, \bibinfo {author}
  {\bibfnamefont {T.~Y.}\ \bibnamefont {Wu}}, \bibinfo {author} {\bibfnamefont
  {X.}~\bibnamefont {Xie}},\ and\ \bibinfo {author} {\bibfnamefont {B.~J.}\
  \bibnamefont {Bloom}},\ }\bibfield  {title} {\bibinfo {title} {{High-fidelity
  universal gates in the $^{171}$Yb ground state nuclear spin qubit}},\ }\href
  {http://arxiv.org/abs/2411.11708} {\bibfield  {journal} {\bibinfo  {journal}
  {arXiv Prepr.}\ }\textbf {\bibinfo {volume} {2411.11708}} (\bibinfo {year}
  {2024})}\BibitemShut {NoStop}%
\bibitem [{\citenamefont {Peper}\ \emph {et~al.}(2024)\citenamefont {Peper},
  \citenamefont {Li}, \citenamefont {Knapp}, \citenamefont {Bileska},
  \citenamefont {Ma}, \citenamefont {Liu}, \citenamefont {Peng}, \citenamefont
  {Zhang}, \citenamefont {Horvath}, \citenamefont {Burgers},\ and\
  \citenamefont {Thompson}}]{Peper2024}%
  \BibitemOpen
  \bibfield  {author} {\bibinfo {author} {\bibfnamefont {M.}~\bibnamefont
  {Peper}}, \bibinfo {author} {\bibfnamefont {Y.}~\bibnamefont {Li}}, \bibinfo
  {author} {\bibfnamefont {D.~Y.}\ \bibnamefont {Knapp}}, \bibinfo {author}
  {\bibfnamefont {M.}~\bibnamefont {Bileska}}, \bibinfo {author} {\bibfnamefont
  {S.}~\bibnamefont {Ma}}, \bibinfo {author} {\bibfnamefont {G.}~\bibnamefont
  {Liu}}, \bibinfo {author} {\bibfnamefont {P.}~\bibnamefont {Peng}}, \bibinfo
  {author} {\bibfnamefont {B.}~\bibnamefont {Zhang}}, \bibinfo {author}
  {\bibfnamefont {S.~P.}\ \bibnamefont {Horvath}}, \bibinfo {author}
  {\bibfnamefont {A.~P.}\ \bibnamefont {Burgers}},\ and\ \bibinfo {author}
  {\bibfnamefont {J.~D.}\ \bibnamefont {Thompson}},\ }\bibfield  {title}
  {\bibinfo {title} {{Spectroscopy and modeling of $^{171}$Yb Rydberg states
  for high-fidelity two-qubit gates}},\ }\href
  {http://arxiv.org/abs/2406.01482} {\bibfield  {journal} {\bibinfo  {journal}
  {arXiv Prepr.}\ }\textbf {\bibinfo {volume} {2406.01482}} (\bibinfo {year}
  {2024})}\BibitemShut {NoStop}%
\bibitem [{\citenamefont {Finkelstein}\ \emph {et~al.}(2024)\citenamefont
  {Finkelstein}, \citenamefont {Tsai}, \citenamefont {Sun}, \citenamefont
  {Scholl}, \citenamefont {Direkci}, \citenamefont {Gefen}, \citenamefont
  {Choi}, \citenamefont {Shaw},\ and\ \citenamefont
  {Endres}}]{Finkelstein2024}%
  \BibitemOpen
  \bibfield  {author} {\bibinfo {author} {\bibfnamefont {R.}~\bibnamefont
  {Finkelstein}}, \bibinfo {author} {\bibfnamefont {R.~B.-S.}\ \bibnamefont
  {Tsai}}, \bibinfo {author} {\bibfnamefont {X.}~\bibnamefont {Sun}}, \bibinfo
  {author} {\bibfnamefont {P.}~\bibnamefont {Scholl}}, \bibinfo {author}
  {\bibfnamefont {S.}~\bibnamefont {Direkci}}, \bibinfo {author} {\bibfnamefont
  {T.}~\bibnamefont {Gefen}}, \bibinfo {author} {\bibfnamefont
  {J.}~\bibnamefont {Choi}}, \bibinfo {author} {\bibfnamefont {A.~L.}\
  \bibnamefont {Shaw}},\ and\ \bibinfo {author} {\bibfnamefont
  {M.}~\bibnamefont {Endres}},\ }\bibfield  {title} {\bibinfo {title}
  {{Universal quantum operations and ancilla-based read-out for tweezer
  clocks}},\ }\href {https://doi.org/10.1038/s41586-024-08005-8} {\bibfield
  {journal} {\bibinfo  {journal} {Nature}\ }\textbf {\bibinfo {volume} {634}},\
  \bibinfo {pages} {321} (\bibinfo {year} {2024})}\BibitemShut {NoStop}%
\bibitem [{\citenamefont {Young}\ \emph {et~al.}(2020)\citenamefont {Young},
  \citenamefont {Eckner}, \citenamefont {Milner}, \citenamefont {Kedar},
  \citenamefont {Norcia}, \citenamefont {Oelker}, \citenamefont {Schine},
  \citenamefont {Ye},\ and\ \citenamefont {Kaufman}}]{Young2020}%
  \BibitemOpen
  \bibfield  {author} {\bibinfo {author} {\bibfnamefont {A.~W.}\ \bibnamefont
  {Young}}, \bibinfo {author} {\bibfnamefont {W.~J.}\ \bibnamefont {Eckner}},
  \bibinfo {author} {\bibfnamefont {W.~R.}\ \bibnamefont {Milner}}, \bibinfo
  {author} {\bibfnamefont {D.}~\bibnamefont {Kedar}}, \bibinfo {author}
  {\bibfnamefont {M.~A.}\ \bibnamefont {Norcia}}, \bibinfo {author}
  {\bibfnamefont {E.}~\bibnamefont {Oelker}}, \bibinfo {author} {\bibfnamefont
  {N.}~\bibnamefont {Schine}}, \bibinfo {author} {\bibfnamefont
  {J.}~\bibnamefont {Ye}},\ and\ \bibinfo {author} {\bibfnamefont {A.~M.}\
  \bibnamefont {Kaufman}},\ }\bibfield  {title} {\bibinfo {title}
  {{Half-minute-scale atomic coherence and high relative stability in a tweezer
  clock}},\ }\href {https://doi.org/10.1038/s41586-020-3009-y} {\bibfield
  {journal} {\bibinfo  {journal} {Nature}\ }\textbf {\bibinfo {volume} {588}},\
  \bibinfo {pages} {408} (\bibinfo {year} {2020})}\BibitemShut {NoStop}%
\bibitem [{\citenamefont {Bothwell}\ \emph {et~al.}(2022)\citenamefont
  {Bothwell}, \citenamefont {Kennedy}, \citenamefont {Aeppli}, \citenamefont
  {Kedar}, \citenamefont {Robinson}, \citenamefont {Oelker}, \citenamefont
  {Staron},\ and\ \citenamefont {Ye}}]{Bothwell2022}%
  \BibitemOpen
  \bibfield  {author} {\bibinfo {author} {\bibfnamefont {T.}~\bibnamefont
  {Bothwell}}, \bibinfo {author} {\bibfnamefont {C.~J.}\ \bibnamefont
  {Kennedy}}, \bibinfo {author} {\bibfnamefont {A.}~\bibnamefont {Aeppli}},
  \bibinfo {author} {\bibfnamefont {D.}~\bibnamefont {Kedar}}, \bibinfo
  {author} {\bibfnamefont {J.~M.}\ \bibnamefont {Robinson}}, \bibinfo {author}
  {\bibfnamefont {E.}~\bibnamefont {Oelker}}, \bibinfo {author} {\bibfnamefont
  {A.}~\bibnamefont {Staron}},\ and\ \bibinfo {author} {\bibfnamefont
  {J.}~\bibnamefont {Ye}},\ }\bibfield  {title} {\bibinfo {title} {{Resolving
  the gravitational redshift across a millimetre-scale atomic sample}},\ }\href
  {https://doi.org/10.1038/s41586-021-04349-7} {\bibfield  {journal} {\bibinfo
  {journal} {Nature}\ }\textbf {\bibinfo {volume} {602}},\ \bibinfo {pages}
  {420} (\bibinfo {year} {2022})}\BibitemShut {NoStop}%
\bibitem [{\citenamefont {Schine}\ \emph {et~al.}(2022)\citenamefont {Schine},
  \citenamefont {Young}, \citenamefont {Eckner}, \citenamefont {Martin},\ and\
  \citenamefont {Kaufman}}]{schine2022}%
  \BibitemOpen
  \bibfield  {author} {\bibinfo {author} {\bibfnamefont {N.}~\bibnamefont
  {Schine}}, \bibinfo {author} {\bibfnamefont {A.~W.}\ \bibnamefont {Young}},
  \bibinfo {author} {\bibfnamefont {W.~J.}\ \bibnamefont {Eckner}}, \bibinfo
  {author} {\bibfnamefont {M.~J.}\ \bibnamefont {Martin}},\ and\ \bibinfo
  {author} {\bibfnamefont {A.~M.}\ \bibnamefont {Kaufman}},\ }\bibfield
  {title} {\bibinfo {title} {{Long-lived Bell states in an array of optical
  clock qubits}},\ }\href {https://doi.org/10.1038/s41567-022-01678-w}
  {\bibfield  {journal} {\bibinfo  {journal} {Nat. Phys.}\ }\textbf {\bibinfo
  {volume} {18}},\ \bibinfo {pages} {1067} (\bibinfo {year}
  {2022})}\BibitemShut {NoStop}%
\bibitem [{\citenamefont {Cao}\ \emph {et~al.}(2024)\citenamefont {Cao},
  \citenamefont {Eckner}, \citenamefont {{Lukin Yelin}}, \citenamefont {Young},
  \citenamefont {Jandura}, \citenamefont {Yan}, \citenamefont {Kim},
  \citenamefont {Pupillo}, \citenamefont {Ye}, \citenamefont {{Darkwah
  Oppong}},\ and\ \citenamefont {Kaufman}}]{Cao2024}%
  \BibitemOpen
  \bibfield  {author} {\bibinfo {author} {\bibfnamefont {A.}~\bibnamefont
  {Cao}}, \bibinfo {author} {\bibfnamefont {W.~J.}\ \bibnamefont {Eckner}},
  \bibinfo {author} {\bibfnamefont {T.}~\bibnamefont {{Lukin Yelin}}}, \bibinfo
  {author} {\bibfnamefont {A.~W.}\ \bibnamefont {Young}}, \bibinfo {author}
  {\bibfnamefont {S.}~\bibnamefont {Jandura}}, \bibinfo {author} {\bibfnamefont
  {L.}~\bibnamefont {Yan}}, \bibinfo {author} {\bibfnamefont {K.}~\bibnamefont
  {Kim}}, \bibinfo {author} {\bibfnamefont {G.}~\bibnamefont {Pupillo}},
  \bibinfo {author} {\bibfnamefont {J.}~\bibnamefont {Ye}}, \bibinfo {author}
  {\bibfnamefont {N.}~\bibnamefont {{Darkwah Oppong}}},\ and\ \bibinfo {author}
  {\bibfnamefont {A.~M.}\ \bibnamefont {Kaufman}},\ }\bibfield  {title}
  {\bibinfo {title} {{Multi-qubit gates and Schr{\"{o}}dinger cat states in an
  optical clock}},\ }\href {https://doi.org/10.1038/s41586-024-07913-z}
  {\bibfield  {journal} {\bibinfo  {journal} {Nature}\ }\textbf {\bibinfo
  {volume} {634}},\ \bibinfo {pages} {315} (\bibinfo {year}
  {2024})}\BibitemShut {NoStop}%
\bibitem [{\citenamefont {Lemke}(2012)}]{Lemke2012}%
  \BibitemOpen
  \bibfield  {author} {\bibinfo {author} {\bibfnamefont {N.~D.}\ \bibnamefont
  {Lemke}},\ }\emph {\bibinfo {title} {{Optical Lattice Clock with Spin-1/2
  Ytterbium Atoms}}},\ \href@noop {} {\bibinfo {type} {Ph.d. thesis}},\
  \bibinfo  {school} {University of Colorado-Boulder} (\bibinfo {year}
  {2012})\BibitemShut {NoStop}%
\bibitem [{\citenamefont {Chen}\ \emph {et~al.}(2022)\citenamefont {Chen},
  \citenamefont {Li}, \citenamefont {Huie}, \citenamefont {Zhao}, \citenamefont
  {Vetter}, \citenamefont {Greene},\ and\ \citenamefont {Covey}}]{Chen2022}%
  \BibitemOpen
  \bibfield  {author} {\bibinfo {author} {\bibfnamefont {N.}~\bibnamefont
  {Chen}}, \bibinfo {author} {\bibfnamefont {L.}~\bibnamefont {Li}}, \bibinfo
  {author} {\bibfnamefont {W.}~\bibnamefont {Huie}}, \bibinfo {author}
  {\bibfnamefont {M.}~\bibnamefont {Zhao}}, \bibinfo {author} {\bibfnamefont
  {I.}~\bibnamefont {Vetter}}, \bibinfo {author} {\bibfnamefont {C.~H.}\
  \bibnamefont {Greene}},\ and\ \bibinfo {author} {\bibfnamefont {J.~P.}\
  \bibnamefont {Covey}},\ }\bibfield  {title} {\bibinfo {title} {{Analyzing the
  Rydberg-based optical-metastable-ground architecture for $^{171}$Yb nuclear
  spins}},\ }\href {https://doi.org/10.1103/PhysRevA.105.052438} {\bibfield
  {journal} {\bibinfo  {journal} {Phys. Rev. A}\ }\textbf {\bibinfo {volume}
  {105}},\ \bibinfo {pages} {052438} (\bibinfo {year} {2022})}\BibitemShut
  {NoStop}%
\bibitem [{\citenamefont {Huie}\ \emph {et~al.}(2023)\citenamefont {Huie},
  \citenamefont {Li}, \citenamefont {Chen}, \citenamefont {Hu}, \citenamefont
  {Jia}, \citenamefont {Sun},\ and\ \citenamefont {Covey}}]{Huie2023}%
  \BibitemOpen
  \bibfield  {author} {\bibinfo {author} {\bibfnamefont {W.}~\bibnamefont
  {Huie}}, \bibinfo {author} {\bibfnamefont {L.}~\bibnamefont {Li}}, \bibinfo
  {author} {\bibfnamefont {N.}~\bibnamefont {Chen}}, \bibinfo {author}
  {\bibfnamefont {X.}~\bibnamefont {Hu}}, \bibinfo {author} {\bibfnamefont
  {Z.}~\bibnamefont {Jia}}, \bibinfo {author} {\bibfnamefont {W.~K.~C.}\
  \bibnamefont {Sun}},\ and\ \bibinfo {author} {\bibfnamefont {J.~P.}\
  \bibnamefont {Covey}},\ }\bibfield  {title} {\bibinfo {title} {{Repetitive
  Readout and Real-Time Control of Nuclear Spin Qubits in 171Yb Atoms}},\
  }\href {https://doi.org/10.1103/PRXQuantum.4.030337} {\bibfield  {journal}
  {\bibinfo  {journal} {PRX Quantum}\ }\textbf {\bibinfo {volume} {4}},\
  \bibinfo {pages} {030337} (\bibinfo {year} {2023})}\BibitemShut {NoStop}%
\bibitem [{\citenamefont {Norcia}\ \emph {et~al.}(2023)\citenamefont {Norcia},
  \citenamefont {Cairncross}, \citenamefont {Barnes}, \citenamefont
  {Battaglino}, \citenamefont {Brown}, \citenamefont {Brown}, \citenamefont
  {Cassella}, \citenamefont {Chen}, \citenamefont {Coxe}, \citenamefont {Crow},
  \citenamefont {Epstein}, \citenamefont {Griger}, \citenamefont {Jones},
  \citenamefont {Kim}, \citenamefont {Kindem}, \citenamefont {King},
  \citenamefont {Kondov}, \citenamefont {Kotru}, \citenamefont {Lauigan},
  \citenamefont {Li}, \citenamefont {Lu}, \citenamefont {Megidish},
  \citenamefont {Marjanovic}, \citenamefont {McDonald}, \citenamefont
  {Mittiga}, \citenamefont {Muniz}, \citenamefont {Narayanaswami},
  \citenamefont {Nishiguchi}, \citenamefont {Notermans}, \citenamefont {Paule},
  \citenamefont {Pawlak}, \citenamefont {Peng}, \citenamefont {Ryou},
  \citenamefont {Smull}, \citenamefont {Stack}, \citenamefont {Stone},
  \citenamefont {Sucich}, \citenamefont {Urbanek}, \citenamefont {van~de
  Veerdonk}, \citenamefont {Vendeiro}, \citenamefont {Wilkason}, \citenamefont
  {Wu}, \citenamefont {Xie}, \citenamefont {Zhang},\ and\ \citenamefont
  {Bloom}}]{Norcia2023}%
  \BibitemOpen
  \bibfield  {author} {\bibinfo {author} {\bibfnamefont {M.~A.}\ \bibnamefont
  {Norcia}}, \bibinfo {author} {\bibfnamefont {W.~B.}\ \bibnamefont
  {Cairncross}}, \bibinfo {author} {\bibfnamefont {K.}~\bibnamefont {Barnes}},
  \bibinfo {author} {\bibfnamefont {P.}~\bibnamefont {Battaglino}}, \bibinfo
  {author} {\bibfnamefont {A.}~\bibnamefont {Brown}}, \bibinfo {author}
  {\bibfnamefont {M.~O.}\ \bibnamefont {Brown}}, \bibinfo {author}
  {\bibfnamefont {K.}~\bibnamefont {Cassella}}, \bibinfo {author}
  {\bibfnamefont {C.-A.}\ \bibnamefont {Chen}}, \bibinfo {author}
  {\bibfnamefont {R.}~\bibnamefont {Coxe}}, \bibinfo {author} {\bibfnamefont
  {D.}~\bibnamefont {Crow}}, \bibinfo {author} {\bibfnamefont {J.}~\bibnamefont
  {Epstein}}, \bibinfo {author} {\bibfnamefont {C.}~\bibnamefont {Griger}},
  \bibinfo {author} {\bibfnamefont {A.~M.~W.}\ \bibnamefont {Jones}}, \bibinfo
  {author} {\bibfnamefont {H.}~\bibnamefont {Kim}}, \bibinfo {author}
  {\bibfnamefont {J.~M.}\ \bibnamefont {Kindem}}, \bibinfo {author}
  {\bibfnamefont {J.}~\bibnamefont {King}}, \bibinfo {author} {\bibfnamefont
  {S.~S.}\ \bibnamefont {Kondov}}, \bibinfo {author} {\bibfnamefont
  {K.}~\bibnamefont {Kotru}}, \bibinfo {author} {\bibfnamefont
  {J.}~\bibnamefont {Lauigan}}, \bibinfo {author} {\bibfnamefont
  {M.}~\bibnamefont {Li}}, \bibinfo {author} {\bibfnamefont {M.}~\bibnamefont
  {Lu}}, \bibinfo {author} {\bibfnamefont {E.}~\bibnamefont {Megidish}},
  \bibinfo {author} {\bibfnamefont {J.}~\bibnamefont {Marjanovic}}, \bibinfo
  {author} {\bibfnamefont {M.}~\bibnamefont {McDonald}}, \bibinfo {author}
  {\bibfnamefont {T.}~\bibnamefont {Mittiga}}, \bibinfo {author} {\bibfnamefont
  {J.~A.}\ \bibnamefont {Muniz}}, \bibinfo {author} {\bibfnamefont
  {S.}~\bibnamefont {Narayanaswami}}, \bibinfo {author} {\bibfnamefont
  {C.}~\bibnamefont {Nishiguchi}}, \bibinfo {author} {\bibfnamefont
  {R.}~\bibnamefont {Notermans}}, \bibinfo {author} {\bibfnamefont
  {T.}~\bibnamefont {Paule}}, \bibinfo {author} {\bibfnamefont {K.~A.}\
  \bibnamefont {Pawlak}}, \bibinfo {author} {\bibfnamefont {L.~S.}\
  \bibnamefont {Peng}}, \bibinfo {author} {\bibfnamefont {A.}~\bibnamefont
  {Ryou}}, \bibinfo {author} {\bibfnamefont {A.}~\bibnamefont {Smull}},
  \bibinfo {author} {\bibfnamefont {D.}~\bibnamefont {Stack}}, \bibinfo
  {author} {\bibfnamefont {M.}~\bibnamefont {Stone}}, \bibinfo {author}
  {\bibfnamefont {A.}~\bibnamefont {Sucich}}, \bibinfo {author} {\bibfnamefont
  {M.}~\bibnamefont {Urbanek}}, \bibinfo {author} {\bibfnamefont {R.~J.~M.}\
  \bibnamefont {van~de Veerdonk}}, \bibinfo {author} {\bibfnamefont
  {Z.}~\bibnamefont {Vendeiro}}, \bibinfo {author} {\bibfnamefont
  {T.}~\bibnamefont {Wilkason}}, \bibinfo {author} {\bibfnamefont {T.-Y.}\
  \bibnamefont {Wu}}, \bibinfo {author} {\bibfnamefont {X.}~\bibnamefont
  {Xie}}, \bibinfo {author} {\bibfnamefont {X.}~\bibnamefont {Zhang}},\ and\
  \bibinfo {author} {\bibfnamefont {B.~J.}\ \bibnamefont {Bloom}},\ }\bibfield
  {title} {\bibinfo {title} {{Midcircuit Qubit Measurement and Rearrangement in
  a Yb-171 Atomic Array}},\ }\href {https://doi.org/10.1103/PhysRevX.13.041034}
  {\bibfield  {journal} {\bibinfo  {journal} {Phys. Rev. X}\ }\textbf {\bibinfo
  {volume} {13}},\ \bibinfo {pages} {041034} (\bibinfo {year}
  {2023})}\BibitemShut {NoStop}%
\bibitem [{\citenamefont {Jia}\ \emph {et~al.}(2024)\citenamefont {Jia},
  \citenamefont {Huie}, \citenamefont {Li}, \citenamefont {Sun}, \citenamefont
  {Hu}, \citenamefont {Aakash}, \citenamefont {Kogan}, \citenamefont {Karve},
  \citenamefont {Lee},\ and\ \citenamefont {Covey}}]{Jia2024}%
  \BibitemOpen
  \bibfield  {author} {\bibinfo {author} {\bibfnamefont {Z.}~\bibnamefont
  {Jia}}, \bibinfo {author} {\bibfnamefont {W.}~\bibnamefont {Huie}}, \bibinfo
  {author} {\bibfnamefont {L.}~\bibnamefont {Li}}, \bibinfo {author}
  {\bibfnamefont {W.~K.~C.}\ \bibnamefont {Sun}}, \bibinfo {author}
  {\bibfnamefont {X.}~\bibnamefont {Hu}}, \bibinfo {author} {\bibnamefont
  {Aakash}}, \bibinfo {author} {\bibfnamefont {H.}~\bibnamefont {Kogan}},
  \bibinfo {author} {\bibfnamefont {A.}~\bibnamefont {Karve}}, \bibinfo
  {author} {\bibfnamefont {J.~Y.}\ \bibnamefont {Lee}},\ and\ \bibinfo {author}
  {\bibfnamefont {J.~P.}\ \bibnamefont {Covey}},\ }\bibfield  {title} {\bibinfo
  {title} {{An architecture for two-qubit encoding in neutral ytterbium-171
  atoms}},\ }\href {https://doi.org/10.1038/s41534-024-00898-7} {\bibfield
  {journal} {\bibinfo  {journal} {npj Quantum Inf.}\ }\textbf {\bibinfo
  {volume} {10}},\ \bibinfo {pages} {106} (\bibinfo {year} {2024})}\BibitemShut
  {NoStop}%
\bibitem [{\citenamefont {Ma}\ \emph {et~al.}(1994)\citenamefont {Ma},
  \citenamefont {Jungner}, \citenamefont {Ye},\ and\ \citenamefont
  {Hall}}]{Ma1994}%
  \BibitemOpen
  \bibfield  {author} {\bibinfo {author} {\bibfnamefont {L.-S.}\ \bibnamefont
  {Ma}}, \bibinfo {author} {\bibfnamefont {P.}~\bibnamefont {Jungner}},
  \bibinfo {author} {\bibfnamefont {J.}~\bibnamefont {Ye}},\ and\ \bibinfo
  {author} {\bibfnamefont {J.~L.}\ \bibnamefont {Hall}},\ }\bibfield  {title}
  {\bibinfo {title} {{Delivering the same optical frequency at two places:
  accurate cancellation of phase noise introduced by an optical fiber or other
  time-varying path}},\ }\href {https://doi.org/10.1364/OL.19.001777}
  {\bibfield  {journal} {\bibinfo  {journal} {Opt. Lett.}\ }\textbf {\bibinfo
  {volume} {19}},\ \bibinfo {pages} {1777} (\bibinfo {year}
  {1994})}\BibitemShut {NoStop}%
\bibitem [{\citenamefont {Pizzocaro}\ \emph {et~al.}(2021)\citenamefont
  {Pizzocaro}, \citenamefont {Sekido}, \citenamefont {Takefuji}, \citenamefont
  {Ujihara}, \citenamefont {Hachisu}, \citenamefont {Nemitz}, \citenamefont
  {Tsutsumi}, \citenamefont {Kondo}, \citenamefont {Kawai}, \citenamefont
  {Ichikawa}, \citenamefont {Namba}, \citenamefont {Okamoto}, \citenamefont
  {Takahashi}, \citenamefont {Komuro}, \citenamefont {Clivati}, \citenamefont
  {Bregolin}, \citenamefont {Barbieri}, \citenamefont {Mura}, \citenamefont
  {Cantoni}, \citenamefont {Cerretto}, \citenamefont {Levi}, \citenamefont
  {Maccaferri}, \citenamefont {Roma}, \citenamefont {Bortolotti}, \citenamefont
  {Negusini}, \citenamefont {Ricci}, \citenamefont {Zacchiroli}, \citenamefont
  {Roda}, \citenamefont {Leute}, \citenamefont {Petit}, \citenamefont {Perini},
  \citenamefont {Calonico},\ and\ \citenamefont {Ido}}]{Pizzocaro2021}%
  \BibitemOpen
  \bibfield  {author} {\bibinfo {author} {\bibfnamefont {M.}~\bibnamefont
  {Pizzocaro}}, \bibinfo {author} {\bibfnamefont {M.}~\bibnamefont {Sekido}},
  \bibinfo {author} {\bibfnamefont {K.}~\bibnamefont {Takefuji}}, \bibinfo
  {author} {\bibfnamefont {H.}~\bibnamefont {Ujihara}}, \bibinfo {author}
  {\bibfnamefont {H.}~\bibnamefont {Hachisu}}, \bibinfo {author} {\bibfnamefont
  {N.}~\bibnamefont {Nemitz}}, \bibinfo {author} {\bibfnamefont
  {M.}~\bibnamefont {Tsutsumi}}, \bibinfo {author} {\bibfnamefont
  {T.}~\bibnamefont {Kondo}}, \bibinfo {author} {\bibfnamefont
  {E.}~\bibnamefont {Kawai}}, \bibinfo {author} {\bibfnamefont
  {R.}~\bibnamefont {Ichikawa}}, \bibinfo {author} {\bibfnamefont
  {K.}~\bibnamefont {Namba}}, \bibinfo {author} {\bibfnamefont
  {Y.}~\bibnamefont {Okamoto}}, \bibinfo {author} {\bibfnamefont
  {R.}~\bibnamefont {Takahashi}}, \bibinfo {author} {\bibfnamefont
  {J.}~\bibnamefont {Komuro}}, \bibinfo {author} {\bibfnamefont
  {C.}~\bibnamefont {Clivati}}, \bibinfo {author} {\bibfnamefont
  {F.}~\bibnamefont {Bregolin}}, \bibinfo {author} {\bibfnamefont
  {P.}~\bibnamefont {Barbieri}}, \bibinfo {author} {\bibfnamefont
  {A.}~\bibnamefont {Mura}}, \bibinfo {author} {\bibfnamefont {E.}~\bibnamefont
  {Cantoni}}, \bibinfo {author} {\bibfnamefont {G.}~\bibnamefont {Cerretto}},
  \bibinfo {author} {\bibfnamefont {F.}~\bibnamefont {Levi}}, \bibinfo {author}
  {\bibfnamefont {G.}~\bibnamefont {Maccaferri}}, \bibinfo {author}
  {\bibfnamefont {M.}~\bibnamefont {Roma}}, \bibinfo {author} {\bibfnamefont
  {C.}~\bibnamefont {Bortolotti}}, \bibinfo {author} {\bibfnamefont
  {M.}~\bibnamefont {Negusini}}, \bibinfo {author} {\bibfnamefont
  {R.}~\bibnamefont {Ricci}}, \bibinfo {author} {\bibfnamefont
  {G.}~\bibnamefont {Zacchiroli}}, \bibinfo {author} {\bibfnamefont
  {J.}~\bibnamefont {Roda}}, \bibinfo {author} {\bibfnamefont {J.}~\bibnamefont
  {Leute}}, \bibinfo {author} {\bibfnamefont {G.}~\bibnamefont {Petit}},
  \bibinfo {author} {\bibfnamefont {F.}~\bibnamefont {Perini}}, \bibinfo
  {author} {\bibfnamefont {D.}~\bibnamefont {Calonico}},\ and\ \bibinfo
  {author} {\bibfnamefont {T.}~\bibnamefont {Ido}},\ }\bibfield  {title}
  {\bibinfo {title} {{Intercontinental comparison of optical atomic clocks
  through very long baseline interferometry}},\ }\href
  {https://doi.org/10.1038/s41567-020-01038-6} {\bibfield  {journal} {\bibinfo
  {journal} {Nat. Phys.}\ }\textbf {\bibinfo {volume} {17}},\ \bibinfo {pages}
  {223} (\bibinfo {year} {2021})}\BibitemShut {NoStop}%
\bibitem [{\citenamefont {Gozzard}\ \emph {et~al.}(2022)\citenamefont
  {Gozzard}, \citenamefont {Howard}, \citenamefont {Dix-Matthews},
  \citenamefont {Karpathakis}, \citenamefont {Gravestock},\ and\ \citenamefont
  {Schediwy}}]{Gozzard2022}%
  \BibitemOpen
  \bibfield  {author} {\bibinfo {author} {\bibfnamefont {D.~R.}\ \bibnamefont
  {Gozzard}}, \bibinfo {author} {\bibfnamefont {L.~A.}\ \bibnamefont {Howard}},
  \bibinfo {author} {\bibfnamefont {B.~P.}\ \bibnamefont {Dix-Matthews}},
  \bibinfo {author} {\bibfnamefont {S.~F.~E.}\ \bibnamefont {Karpathakis}},
  \bibinfo {author} {\bibfnamefont {C.~T.}\ \bibnamefont {Gravestock}},\ and\
  \bibinfo {author} {\bibfnamefont {S.~W.}\ \bibnamefont {Schediwy}},\
  }\bibfield  {title} {\bibinfo {title} {{Ultrastable Free-Space Laser Links
  for a Global Network of Optical Atomic Clocks}},\ }\href
  {https://doi.org/10.1103/PhysRevLett.128.020801} {\bibfield  {journal}
  {\bibinfo  {journal} {Phys. Rev. Lett.}\ }\textbf {\bibinfo {volume} {128}},\
  \bibinfo {pages} {020801} (\bibinfo {year} {2022})}\BibitemShut {NoStop}%
\bibitem [{\citenamefont {Ludlow}\ \emph {et~al.}(2015)\citenamefont {Ludlow},
  \citenamefont {Boyd}, \citenamefont {Ye}, \citenamefont {Peik},\ and\
  \citenamefont {Schmidt}}]{Ludlow2015}%
  \BibitemOpen
  \bibfield  {author} {\bibinfo {author} {\bibfnamefont {A.~D.}\ \bibnamefont
  {Ludlow}}, \bibinfo {author} {\bibfnamefont {M.~M.}\ \bibnamefont {Boyd}},
  \bibinfo {author} {\bibfnamefont {J.}~\bibnamefont {Ye}}, \bibinfo {author}
  {\bibfnamefont {E.}~\bibnamefont {Peik}},\ and\ \bibinfo {author}
  {\bibfnamefont {P.~O.}\ \bibnamefont {Schmidt}},\ }\bibfield  {title}
  {\bibinfo {title} {{Optical atomic clocks}},\ }\href
  {https://doi.org/10.1103/RevModPhys.87.637} {\bibfield  {journal} {\bibinfo
  {journal} {Rev. Mod. Phys.}\ }\textbf {\bibinfo {volume} {87}},\ \bibinfo
  {pages} {637} (\bibinfo {year} {2015})}\BibitemShut {NoStop}%
\bibitem [{\citenamefont {Wcis{\l}o}\ \emph {et~al.}(2018)\citenamefont
  {Wcis{\l}o}, \citenamefont {Ablewski}, \citenamefont {Beloy}, \citenamefont
  {Bilicki}, \citenamefont {Bober}, \citenamefont {Brown}, \citenamefont
  {Fasano}, \citenamefont {Ciury{\l}o}, \citenamefont {Hachisu}, \citenamefont
  {Ido}, \citenamefont {Lodewyck}, \citenamefont {Ludlow}, \citenamefont
  {McGrew}, \citenamefont {Morzy{\'{n}}ski}, \citenamefont {Nicolodi},
  \citenamefont {Schioppo}, \citenamefont {Sekido}, \citenamefont {{Le
  Targat}}, \citenamefont {Wolf}, \citenamefont {Zhang}, \citenamefont
  {Zjawin},\ and\ \citenamefont {Zawada}}]{Wcislo2018}%
  \BibitemOpen
  \bibfield  {author} {\bibinfo {author} {\bibfnamefont {P.}~\bibnamefont
  {Wcis{\l}o}}, \bibinfo {author} {\bibfnamefont {P.}~\bibnamefont {Ablewski}},
  \bibinfo {author} {\bibfnamefont {K.}~\bibnamefont {Beloy}}, \bibinfo
  {author} {\bibfnamefont {S.}~\bibnamefont {Bilicki}}, \bibinfo {author}
  {\bibfnamefont {M.}~\bibnamefont {Bober}}, \bibinfo {author} {\bibfnamefont
  {R.}~\bibnamefont {Brown}}, \bibinfo {author} {\bibfnamefont
  {R.}~\bibnamefont {Fasano}}, \bibinfo {author} {\bibfnamefont
  {R.}~\bibnamefont {Ciury{\l}o}}, \bibinfo {author} {\bibfnamefont
  {H.}~\bibnamefont {Hachisu}}, \bibinfo {author} {\bibfnamefont
  {T.}~\bibnamefont {Ido}}, \bibinfo {author} {\bibfnamefont {J.}~\bibnamefont
  {Lodewyck}}, \bibinfo {author} {\bibfnamefont {A.}~\bibnamefont {Ludlow}},
  \bibinfo {author} {\bibfnamefont {W.}~\bibnamefont {McGrew}}, \bibinfo
  {author} {\bibfnamefont {P.}~\bibnamefont {Morzy{\'{n}}ski}}, \bibinfo
  {author} {\bibfnamefont {D.}~\bibnamefont {Nicolodi}}, \bibinfo {author}
  {\bibfnamefont {M.}~\bibnamefont {Schioppo}}, \bibinfo {author}
  {\bibfnamefont {M.}~\bibnamefont {Sekido}}, \bibinfo {author} {\bibfnamefont
  {R.}~\bibnamefont {{Le Targat}}}, \bibinfo {author} {\bibfnamefont
  {P.}~\bibnamefont {Wolf}}, \bibinfo {author} {\bibfnamefont {X.}~\bibnamefont
  {Zhang}}, \bibinfo {author} {\bibfnamefont {B.}~\bibnamefont {Zjawin}},\ and\
  \bibinfo {author} {\bibfnamefont {M.}~\bibnamefont {Zawada}},\ }\bibfield
  {title} {\bibinfo {title} {{New bounds on dark matter coupling from a global
  network of optical atomic clocks}},\ }\bibfield  {journal} {\bibinfo
  {journal} {Sci. Adv.}\ }\textbf {\bibinfo {volume} {4}},\ \href
  {https://doi.org/10.1126/sciadv.aau4869} {10.1126/sciadv.aau4869} (\bibinfo
  {year} {2018})\BibitemShut {NoStop}%
\bibitem [{\citenamefont {Singh}\ \emph {et~al.}(2023)\citenamefont {Singh},
  \citenamefont {Bradley}, \citenamefont {Anand}, \citenamefont {Ramesh},
  \citenamefont {White},\ and\ \citenamefont {Bernien}}]{Singh2022}%
  \BibitemOpen
  \bibfield  {author} {\bibinfo {author} {\bibfnamefont {K.}~\bibnamefont
  {Singh}}, \bibinfo {author} {\bibfnamefont {C.~E.}\ \bibnamefont {Bradley}},
  \bibinfo {author} {\bibfnamefont {S.}~\bibnamefont {Anand}}, \bibinfo
  {author} {\bibfnamefont {V.}~\bibnamefont {Ramesh}}, \bibinfo {author}
  {\bibfnamefont {R.}~\bibnamefont {White}},\ and\ \bibinfo {author}
  {\bibfnamefont {H.}~\bibnamefont {Bernien}},\ }\bibfield  {title} {\bibinfo
  {title} {{Mid-circuit correction of correlated phase errors using an array of
  spectator qubits}},\ }\href {https://doi.org/10.1126/science.ade5337}
  {\bibfield  {journal} {\bibinfo  {journal} {Science}\ }\textbf {\bibinfo
  {volume} {380}},\ \bibinfo {pages} {1265} (\bibinfo {year}
  {2023})}\BibitemShut {NoStop}%
\bibitem [{\citenamefont {Asenbaum}\ \emph {et~al.}(2017)\citenamefont
  {Asenbaum}, \citenamefont {Overstreet}, \citenamefont {Kovachy},
  \citenamefont {Brown}, \citenamefont {Hogan},\ and\ \citenamefont
  {Kasevich}}]{asenbaum2017phase}%
  \BibitemOpen
  \bibfield  {author} {\bibinfo {author} {\bibfnamefont {P.}~\bibnamefont
  {Asenbaum}}, \bibinfo {author} {\bibfnamefont {C.}~\bibnamefont
  {Overstreet}}, \bibinfo {author} {\bibfnamefont {T.}~\bibnamefont {Kovachy}},
  \bibinfo {author} {\bibfnamefont {D.~D.}\ \bibnamefont {Brown}}, \bibinfo
  {author} {\bibfnamefont {J.~M.}\ \bibnamefont {Hogan}},\ and\ \bibinfo
  {author} {\bibfnamefont {M.~A.}\ \bibnamefont {Kasevich}},\ }\bibfield
  {title} {\bibinfo {title} {Phase shift in an atom interferometer due to
  spacetime curvature across its wave function},\ }\href@noop {} {\bibfield
  {journal} {\bibinfo  {journal} {Physical review letters}\ }\textbf {\bibinfo
  {volume} {118}},\ \bibinfo {pages} {183602} (\bibinfo {year}
  {2017})}\BibitemShut {NoStop}%
\bibitem [{\citenamefont {Penrose}(1996)}]{penrose1996gravity}%
  \BibitemOpen
  \bibfield  {author} {\bibinfo {author} {\bibfnamefont {R.}~\bibnamefont
  {Penrose}},\ }\bibfield  {title} {\bibinfo {title} {On gravity's role in
  quantum state reduction},\ }\href {https://doi.org/10.1007/BF02105068}
  {\bibfield  {journal} {\bibinfo  {journal} {General relativity and
  gravitation}\ }\textbf {\bibinfo {volume} {28}},\ \bibinfo {pages} {581}
  (\bibinfo {year} {1996})}\BibitemShut {NoStop}%
\bibitem [{\citenamefont {Bekenstein}(2014)}]{bekenstein2014can}%
  \BibitemOpen
  \bibfield  {author} {\bibinfo {author} {\bibfnamefont {J.~D.}\ \bibnamefont
  {Bekenstein}},\ }\bibfield  {title} {\bibinfo {title} {Can quantum gravity be
  exposed in the laboratory?},\ }\href
  {https://doi.org/10.1007/s10701-014-9779-z} {\bibfield  {journal} {\bibinfo
  {journal} {Foundations of Physics}\ }\textbf {\bibinfo {volume} {44}},\
  \bibinfo {pages} {452} (\bibinfo {year} {2014})}\BibitemShut {NoStop}%
\bibitem [{\citenamefont {Ralph}\ and\ \citenamefont
  {Pienaar}(2014)}]{ralph2014entanglement}%
  \BibitemOpen
  \bibfield  {author} {\bibinfo {author} {\bibfnamefont {T.}~\bibnamefont
  {Ralph}}\ and\ \bibinfo {author} {\bibfnamefont {J.}~\bibnamefont
  {Pienaar}},\ }\bibfield  {title} {\bibinfo {title} {Entanglement decoherence
  in a gravitational well according to the event formalism},\ }\href
  {https://doi.org/10.1088/1367-2630/16/8/085008} {\bibfield  {journal}
  {\bibinfo  {journal} {New Journal of Physics}\ }\textbf {\bibinfo {volume}
  {16}},\ \bibinfo {pages} {085008} (\bibinfo {year} {2014})}\BibitemShut
  {NoStop}%
\bibitem [{\citenamefont {Rydving}\ \emph {et~al.}(2021)\citenamefont
  {Rydving}, \citenamefont {Aurell},\ and\ \citenamefont
  {Pikovski}}]{rydving2021gedanken}%
  \BibitemOpen
  \bibfield  {author} {\bibinfo {author} {\bibfnamefont {E.}~\bibnamefont
  {Rydving}}, \bibinfo {author} {\bibfnamefont {E.}~\bibnamefont {Aurell}},\
  and\ \bibinfo {author} {\bibfnamefont {I.}~\bibnamefont {Pikovski}},\
  }\bibfield  {title} {\bibinfo {title} {Do gedanken experiments compel
  quantization of gravity?},\ }\href
  {https://doi.org/10.1103/PhysRevD.104.086024} {\bibfield  {journal} {\bibinfo
   {journal} {Physical Review D}\ }\textbf {\bibinfo {volume} {104}},\ \bibinfo
  {pages} {086024} (\bibinfo {year} {2021})}\BibitemShut {NoStop}%
\bibitem [{\citenamefont {Xu}\ \emph {et~al.}(2019)\citenamefont {Xu},
  \citenamefont {Ma}, \citenamefont {Ren}, \citenamefont {Yong}, \citenamefont
  {Ralph}, \citenamefont {Liao}, \citenamefont {Yin}, \citenamefont {Liu},
  \citenamefont {Cai}, \citenamefont {Han} \emph {et~al.}}]{xu2019satellite}%
  \BibitemOpen
  \bibfield  {author} {\bibinfo {author} {\bibfnamefont {P.}~\bibnamefont
  {Xu}}, \bibinfo {author} {\bibfnamefont {Y.}~\bibnamefont {Ma}}, \bibinfo
  {author} {\bibfnamefont {J.-G.}\ \bibnamefont {Ren}}, \bibinfo {author}
  {\bibfnamefont {H.-L.}\ \bibnamefont {Yong}}, \bibinfo {author}
  {\bibfnamefont {T.~C.}\ \bibnamefont {Ralph}}, \bibinfo {author}
  {\bibfnamefont {S.-K.}\ \bibnamefont {Liao}}, \bibinfo {author}
  {\bibfnamefont {J.}~\bibnamefont {Yin}}, \bibinfo {author} {\bibfnamefont
  {W.-Y.}\ \bibnamefont {Liu}}, \bibinfo {author} {\bibfnamefont {W.-Q.}\
  \bibnamefont {Cai}}, \bibinfo {author} {\bibfnamefont {X.}~\bibnamefont
  {Han}}, \emph {et~al.},\ }\bibfield  {title} {\bibinfo {title} {Satellite
  testing of a gravitationally induced quantum decoherence model},\ }\href
  {https://doi.org/10.1126/science.aay582} {\bibfield  {journal} {\bibinfo
  {journal} {Science}\ }\textbf {\bibinfo {volume} {366}},\ \bibinfo {pages}
  {132} (\bibinfo {year} {2019})}\BibitemShut {NoStop}%
\bibitem [{\citenamefont {Marchese}\ \emph {et~al.}(2024)\citenamefont
  {Marchese}, \citenamefont {Pl{\'a}vala}, \citenamefont {Kleinmann},\ and\
  \citenamefont {Nimmrichter}}]{marchese2024newton}%
  \BibitemOpen
  \bibfield  {author} {\bibinfo {author} {\bibfnamefont {M.~M.}\ \bibnamefont
  {Marchese}}, \bibinfo {author} {\bibfnamefont {M.}~\bibnamefont
  {Pl{\'a}vala}}, \bibinfo {author} {\bibfnamefont {M.}~\bibnamefont
  {Kleinmann}},\ and\ \bibinfo {author} {\bibfnamefont {S.}~\bibnamefont
  {Nimmrichter}},\ }\bibfield  {title} {\bibinfo {title} {Newton's laws of
  motion can generate gravity-mediated entanglement},\ }\href@noop {}
  {\bibfield  {journal} {\bibinfo  {journal} {arXiv preprint arXiv:2401.07832}\
  } (\bibinfo {year} {2024})}\BibitemShut {NoStop}%
\bibitem [{\citenamefont {Pl{\'a}vala}\ \emph {et~al.}(2025)\citenamefont
  {Pl{\'a}vala}, \citenamefont {Nimmrichter},\ and\ \citenamefont
  {Kleinmann}}]{plavala2025probing}%
  \BibitemOpen
  \bibfield  {author} {\bibinfo {author} {\bibfnamefont {M.}~\bibnamefont
  {Pl{\'a}vala}}, \bibinfo {author} {\bibfnamefont {S.}~\bibnamefont
  {Nimmrichter}},\ and\ \bibinfo {author} {\bibfnamefont {M.}~\bibnamefont
  {Kleinmann}},\ }\bibfield  {title} {\bibinfo {title} {Probing the
  nonclassical dynamics of a quantum particle in a gravitational field},\
  }\bibfield  {journal} {\bibinfo  {journal} {arXiv preprint arXiv:2502.03489}\
  }\href {https://doi.org/10.48550/arXiv.2502.03489}
  {10.48550/arXiv.2502.03489} (\bibinfo {year} {2025})\BibitemShut {NoStop}%
\bibitem [{\citenamefont {Sorkin}(1994)}]{sorkin1994quantum}%
  \BibitemOpen
  \bibfield  {author} {\bibinfo {author} {\bibfnamefont {R.~D.}\ \bibnamefont
  {Sorkin}},\ }\bibfield  {title} {\bibinfo {title} {Quantum mechanics as
  quantum measure theory},\ }\href@noop {} {\bibfield  {journal} {\bibinfo
  {journal} {Modern Physics Letters A}\ }\textbf {\bibinfo {volume} {9}},\
  \bibinfo {pages} {3119} (\bibinfo {year} {1994})}\BibitemShut {NoStop}%
\bibitem [{\citenamefont {Sinha}\ \emph {et~al.}(2010)\citenamefont {Sinha},
  \citenamefont {Couteau}, \citenamefont {Jennewein}, \citenamefont
  {Laflamme},\ and\ \citenamefont {Weihs}}]{sinha2010ruling}%
  \BibitemOpen
  \bibfield  {author} {\bibinfo {author} {\bibfnamefont {U.}~\bibnamefont
  {Sinha}}, \bibinfo {author} {\bibfnamefont {C.}~\bibnamefont {Couteau}},
  \bibinfo {author} {\bibfnamefont {T.}~\bibnamefont {Jennewein}}, \bibinfo
  {author} {\bibfnamefont {R.}~\bibnamefont {Laflamme}},\ and\ \bibinfo
  {author} {\bibfnamefont {G.}~\bibnamefont {Weihs}},\ }\bibfield  {title}
  {\bibinfo {title} {Ruling out multi-order interference in quantum
  mechanics},\ }\href@noop {} {\bibfield  {journal} {\bibinfo  {journal}
  {Science}\ }\textbf {\bibinfo {volume} {329}},\ \bibinfo {pages} {418}
  (\bibinfo {year} {2010})}\BibitemShut {NoStop}%
\bibitem [{\citenamefont {Park}\ \emph {et~al.}(2012)\citenamefont {Park},
  \citenamefont {Moussa},\ and\ \citenamefont {Laflamme}}]{park2012three}%
  \BibitemOpen
  \bibfield  {author} {\bibinfo {author} {\bibfnamefont {D.~K.}\ \bibnamefont
  {Park}}, \bibinfo {author} {\bibfnamefont {O.}~\bibnamefont {Moussa}},\ and\
  \bibinfo {author} {\bibfnamefont {R.}~\bibnamefont {Laflamme}},\ }\bibfield
  {title} {\bibinfo {title} {Three path interference using nuclear magnetic
  resonance: a test of the consistency of born's rule},\ }\href@noop {}
  {\bibfield  {journal} {\bibinfo  {journal} {New Journal of Physics}\ }\textbf
  {\bibinfo {volume} {14}},\ \bibinfo {pages} {113025} (\bibinfo {year}
  {2012})}\BibitemShut {NoStop}%
\bibitem [{\citenamefont {Valentini}(2023)}]{valentini2023beyond}%
  \BibitemOpen
  \bibfield  {author} {\bibinfo {author} {\bibfnamefont {A.}~\bibnamefont
  {Valentini}},\ }\bibfield  {title} {\bibinfo {title} {Beyond the born rule in
  quantum gravity},\ }\href@noop {} {\bibfield  {journal} {\bibinfo  {journal}
  {Foundations of Physics}\ }\textbf {\bibinfo {volume} {53}},\ \bibinfo
  {pages} {6} (\bibinfo {year} {2023})}\BibitemShut {NoStop}%
\bibitem [{\citenamefont {Graham}\ \emph {et~al.}(2022)\citenamefont {Graham},
  \citenamefont {Song}, \citenamefont {Scott}, \citenamefont {Poole},
  \citenamefont {Phuttitarn}, \citenamefont {Jooya}, \citenamefont {Eichler},
  \citenamefont {Jiang}, \citenamefont {Marra}, \citenamefont {Grinkemeyer},
  \citenamefont {Kwon}, \citenamefont {Ebert}, \citenamefont {Cherek},
  \citenamefont {Lichtman}, \citenamefont {Gillette}, \citenamefont {Gilbert},
  \citenamefont {Bowman}, \citenamefont {Ballance}, \citenamefont {Campbell},
  \citenamefont {Dahl}, \citenamefont {Crawford}, \citenamefont {Blunt},
  \citenamefont {Rogers}, \citenamefont {Noel},\ and\ \citenamefont
  {Saffman}}]{Graham2022}%
  \BibitemOpen
  \bibfield  {author} {\bibinfo {author} {\bibfnamefont {T.~M.}\ \bibnamefont
  {Graham}}, \bibinfo {author} {\bibfnamefont {Y.}~\bibnamefont {Song}},
  \bibinfo {author} {\bibfnamefont {J.}~\bibnamefont {Scott}}, \bibinfo
  {author} {\bibfnamefont {C.}~\bibnamefont {Poole}}, \bibinfo {author}
  {\bibfnamefont {L.}~\bibnamefont {Phuttitarn}}, \bibinfo {author}
  {\bibfnamefont {K.}~\bibnamefont {Jooya}}, \bibinfo {author} {\bibfnamefont
  {P.}~\bibnamefont {Eichler}}, \bibinfo {author} {\bibfnamefont
  {X.}~\bibnamefont {Jiang}}, \bibinfo {author} {\bibfnamefont
  {A.}~\bibnamefont {Marra}}, \bibinfo {author} {\bibfnamefont
  {B.}~\bibnamefont {Grinkemeyer}}, \bibinfo {author} {\bibfnamefont
  {M.}~\bibnamefont {Kwon}}, \bibinfo {author} {\bibfnamefont {M.}~\bibnamefont
  {Ebert}}, \bibinfo {author} {\bibfnamefont {J.}~\bibnamefont {Cherek}},
  \bibinfo {author} {\bibfnamefont {M.~T.}\ \bibnamefont {Lichtman}}, \bibinfo
  {author} {\bibfnamefont {M.}~\bibnamefont {Gillette}}, \bibinfo {author}
  {\bibfnamefont {J.}~\bibnamefont {Gilbert}}, \bibinfo {author} {\bibfnamefont
  {D.}~\bibnamefont {Bowman}}, \bibinfo {author} {\bibfnamefont
  {T.}~\bibnamefont {Ballance}}, \bibinfo {author} {\bibfnamefont
  {C.}~\bibnamefont {Campbell}}, \bibinfo {author} {\bibfnamefont {E.~D.}\
  \bibnamefont {Dahl}}, \bibinfo {author} {\bibfnamefont {O.}~\bibnamefont
  {Crawford}}, \bibinfo {author} {\bibfnamefont {N.~S.}\ \bibnamefont {Blunt}},
  \bibinfo {author} {\bibfnamefont {B.}~\bibnamefont {Rogers}}, \bibinfo
  {author} {\bibfnamefont {T.}~\bibnamefont {Noel}},\ and\ \bibinfo {author}
  {\bibfnamefont {M.}~\bibnamefont {Saffman}},\ }\bibfield  {title} {\bibinfo
  {title} {{Multi-qubit entanglement and algorithms on a neutral-atom quantum
  computer}},\ }\href {https://doi.org/10.1038/s41586-022-04603-6} {\bibfield
  {journal} {\bibinfo  {journal} {Nature}\ }\textbf {\bibinfo {volume} {604}},\
  \bibinfo {pages} {457} (\bibinfo {year} {2022})}\BibitemShut {NoStop}%
\bibitem [{\citenamefont {Bluvstein}\ \emph {et~al.}(2022)\citenamefont
  {Bluvstein}, \citenamefont {Levine}, \citenamefont {Semeghini}, \citenamefont
  {Wang}, \citenamefont {Ebadi}, \citenamefont {Kalinowski}, \citenamefont
  {Keesling}, \citenamefont {Maskara}, \citenamefont {Pichler}, \citenamefont
  {Greiner}, \citenamefont {Vuleti{\'{c}}},\ and\ \citenamefont
  {Lukin}}]{Bluvstein2022}%
  \BibitemOpen
  \bibfield  {author} {\bibinfo {author} {\bibfnamefont {D.}~\bibnamefont
  {Bluvstein}}, \bibinfo {author} {\bibfnamefont {H.}~\bibnamefont {Levine}},
  \bibinfo {author} {\bibfnamefont {G.}~\bibnamefont {Semeghini}}, \bibinfo
  {author} {\bibfnamefont {T.~T.}\ \bibnamefont {Wang}}, \bibinfo {author}
  {\bibfnamefont {S.}~\bibnamefont {Ebadi}}, \bibinfo {author} {\bibfnamefont
  {M.}~\bibnamefont {Kalinowski}}, \bibinfo {author} {\bibfnamefont
  {A.}~\bibnamefont {Keesling}}, \bibinfo {author} {\bibfnamefont
  {N.}~\bibnamefont {Maskara}}, \bibinfo {author} {\bibfnamefont
  {H.}~\bibnamefont {Pichler}}, \bibinfo {author} {\bibfnamefont
  {M.}~\bibnamefont {Greiner}}, \bibinfo {author} {\bibfnamefont
  {V.}~\bibnamefont {Vuleti{\'{c}}}},\ and\ \bibinfo {author} {\bibfnamefont
  {M.~D.}\ \bibnamefont {Lukin}},\ }\bibfield  {title} {\bibinfo {title} {{A
  quantum processor based on coherent transport of entangled atom arrays}},\
  }\href {https://doi.org/10.1038/s41586-022-04592-6} {\bibfield  {journal}
  {\bibinfo  {journal} {Nature}\ }\textbf {\bibinfo {volume} {604}},\ \bibinfo
  {pages} {451} (\bibinfo {year} {2022})}\BibitemShut {NoStop}%
\bibitem [{\citenamefont {Ringbauer}\ \emph {et~al.}(2022)\citenamefont
  {Ringbauer}, \citenamefont {Meth}, \citenamefont {Postler}, \citenamefont
  {Stricker}, \citenamefont {Blatt}, \citenamefont {Schindler},\ and\
  \citenamefont {Monz}}]{Ringbauer2022}%
  \BibitemOpen
  \bibfield  {author} {\bibinfo {author} {\bibfnamefont {M.}~\bibnamefont
  {Ringbauer}}, \bibinfo {author} {\bibfnamefont {M.}~\bibnamefont {Meth}},
  \bibinfo {author} {\bibfnamefont {L.}~\bibnamefont {Postler}}, \bibinfo
  {author} {\bibfnamefont {R.}~\bibnamefont {Stricker}}, \bibinfo {author}
  {\bibfnamefont {R.}~\bibnamefont {Blatt}}, \bibinfo {author} {\bibfnamefont
  {P.}~\bibnamefont {Schindler}},\ and\ \bibinfo {author} {\bibfnamefont
  {T.}~\bibnamefont {Monz}},\ }\bibfield  {title} {\bibinfo {title} {{A
  universal qudit quantum processor with trapped ions}},\ }\bibfield  {journal}
  {\bibinfo  {journal} {Nat. Phys.}\ }\href
  {https://doi.org/10.1038/s41567-022-01658-0} {10.1038/s41567-022-01658-0}
  (\bibinfo {year} {2022})\BibitemShut {NoStop}%
\bibitem [{\citenamefont {Evered}\ \emph {et~al.}(2023)\citenamefont {Evered},
  \citenamefont {Bluvstein}, \citenamefont {Kalinowski}, \citenamefont {Ebadi},
  \citenamefont {Manovitz}, \citenamefont {Zhou}, \citenamefont {Li},
  \citenamefont {Geim}, \citenamefont {Wang}, \citenamefont {Maskara},
  \citenamefont {Levine}, \citenamefont {Semeghini}, \citenamefont {Greiner},
  \citenamefont {Vuleti{\'{c}}},\ and\ \citenamefont {Lukin}}]{Evered2023}%
  \BibitemOpen
  \bibfield  {author} {\bibinfo {author} {\bibfnamefont {S.~J.}\ \bibnamefont
  {Evered}}, \bibinfo {author} {\bibfnamefont {D.}~\bibnamefont {Bluvstein}},
  \bibinfo {author} {\bibfnamefont {M.}~\bibnamefont {Kalinowski}}, \bibinfo
  {author} {\bibfnamefont {S.}~\bibnamefont {Ebadi}}, \bibinfo {author}
  {\bibfnamefont {T.}~\bibnamefont {Manovitz}}, \bibinfo {author}
  {\bibfnamefont {H.}~\bibnamefont {Zhou}}, \bibinfo {author} {\bibfnamefont
  {S.~H.}\ \bibnamefont {Li}}, \bibinfo {author} {\bibfnamefont {A.~A.}\
  \bibnamefont {Geim}}, \bibinfo {author} {\bibfnamefont {T.~T.}\ \bibnamefont
  {Wang}}, \bibinfo {author} {\bibfnamefont {N.}~\bibnamefont {Maskara}},
  \bibinfo {author} {\bibfnamefont {H.}~\bibnamefont {Levine}}, \bibinfo
  {author} {\bibfnamefont {G.}~\bibnamefont {Semeghini}}, \bibinfo {author}
  {\bibfnamefont {M.}~\bibnamefont {Greiner}}, \bibinfo {author} {\bibfnamefont
  {V.}~\bibnamefont {Vuleti{\'{c}}}},\ and\ \bibinfo {author} {\bibfnamefont
  {M.~D.}\ \bibnamefont {Lukin}},\ }\bibfield  {title} {\bibinfo {title}
  {{High-fidelity parallel entangling gates on a neutral-atom quantum
  computer}},\ }\href {https://doi.org/10.1038/s41586-023-06481-y} {\bibfield
  {journal} {\bibinfo  {journal} {Nature}\ }\textbf {\bibinfo {volume} {622}},\
  \bibinfo {pages} {268} (\bibinfo {year} {2023})}\BibitemShut {NoStop}%
\bibitem [{\citenamefont {Radnaev}\ \emph {et~al.}(2024)\citenamefont
  {Radnaev}, \citenamefont {Chung}, \citenamefont {Cole}, \citenamefont
  {Mason}, \citenamefont {Ballance}, \citenamefont {Bedalov}, \citenamefont
  {Belknap}, \citenamefont {Berman}, \citenamefont {Blakely}, \citenamefont
  {Bloomfield}, \citenamefont {Buttler}, \citenamefont {Campbell},
  \citenamefont {Chopinaud}, \citenamefont {Copenhaver}, \citenamefont {Dawes},
  \citenamefont {Eubanks}, \citenamefont {Friss}, \citenamefont {Garcia},
  \citenamefont {Gilbert}, \citenamefont {Gillette}, \citenamefont {Goiporia},
  \citenamefont {Gokhale}, \citenamefont {Goldwin}, \citenamefont {Goodwin},
  \citenamefont {Graham}, \citenamefont {Guttormsson}, \citenamefont {Hickman},
  \citenamefont {Hurtley}, \citenamefont {Iliev}, \citenamefont {Jones},
  \citenamefont {Jones}, \citenamefont {Kuper}, \citenamefont {Lewis},
  \citenamefont {Lichtman}, \citenamefont {Majdeteimouri}, \citenamefont
  {Mason}, \citenamefont {McMaster}, \citenamefont {Miles}, \citenamefont
  {Mitchell}, \citenamefont {Murphree}, \citenamefont {Neff-Mallon},
  \citenamefont {Oh}, \citenamefont {Omole}, \citenamefont {Simon},
  \citenamefont {Pederson}, \citenamefont {Perlin}, \citenamefont {Reiter},
  \citenamefont {Rines}, \citenamefont {Romlow}, \citenamefont {Scott},
  \citenamefont {Stiefvater}, \citenamefont {Tanner}, \citenamefont {Tucker},
  \citenamefont {Vinogradov}, \citenamefont {Warter}, \citenamefont {Yeo},
  \citenamefont {Saffman},\ and\ \citenamefont {Noel}}]{Radnaev2024}%
  \BibitemOpen
  \bibfield  {author} {\bibinfo {author} {\bibfnamefont {A.~G.}\ \bibnamefont
  {Radnaev}}, \bibinfo {author} {\bibfnamefont {W.~C.}\ \bibnamefont {Chung}},
  \bibinfo {author} {\bibfnamefont {D.~C.}\ \bibnamefont {Cole}}, \bibinfo
  {author} {\bibfnamefont {D.}~\bibnamefont {Mason}}, \bibinfo {author}
  {\bibfnamefont {T.~G.}\ \bibnamefont {Ballance}}, \bibinfo {author}
  {\bibfnamefont {M.~J.}\ \bibnamefont {Bedalov}}, \bibinfo {author}
  {\bibfnamefont {D.~A.}\ \bibnamefont {Belknap}}, \bibinfo {author}
  {\bibfnamefont {M.~R.}\ \bibnamefont {Berman}}, \bibinfo {author}
  {\bibfnamefont {M.}~\bibnamefont {Blakely}}, \bibinfo {author} {\bibfnamefont
  {I.~L.}\ \bibnamefont {Bloomfield}}, \bibinfo {author} {\bibfnamefont
  {P.~D.}\ \bibnamefont {Buttler}}, \bibinfo {author} {\bibfnamefont
  {C.}~\bibnamefont {Campbell}}, \bibinfo {author} {\bibfnamefont
  {A.}~\bibnamefont {Chopinaud}}, \bibinfo {author} {\bibfnamefont
  {E.}~\bibnamefont {Copenhaver}}, \bibinfo {author} {\bibfnamefont {M.~K.}\
  \bibnamefont {Dawes}}, \bibinfo {author} {\bibfnamefont {S.~Y.}\ \bibnamefont
  {Eubanks}}, \bibinfo {author} {\bibfnamefont {A.~J.}\ \bibnamefont {Friss}},
  \bibinfo {author} {\bibfnamefont {D.~M.}\ \bibnamefont {Garcia}}, \bibinfo
  {author} {\bibfnamefont {J.}~\bibnamefont {Gilbert}}, \bibinfo {author}
  {\bibfnamefont {M.}~\bibnamefont {Gillette}}, \bibinfo {author}
  {\bibfnamefont {P.}~\bibnamefont {Goiporia}}, \bibinfo {author}
  {\bibfnamefont {P.}~\bibnamefont {Gokhale}}, \bibinfo {author} {\bibfnamefont
  {J.}~\bibnamefont {Goldwin}}, \bibinfo {author} {\bibfnamefont
  {D.}~\bibnamefont {Goodwin}}, \bibinfo {author} {\bibfnamefont {T.~M.}\
  \bibnamefont {Graham}}, \bibinfo {author} {\bibfnamefont {C.}~\bibnamefont
  {Guttormsson}}, \bibinfo {author} {\bibfnamefont {G.~T.}\ \bibnamefont
  {Hickman}}, \bibinfo {author} {\bibfnamefont {L.}~\bibnamefont {Hurtley}},
  \bibinfo {author} {\bibfnamefont {M.}~\bibnamefont {Iliev}}, \bibinfo
  {author} {\bibfnamefont {E.~B.}\ \bibnamefont {Jones}}, \bibinfo {author}
  {\bibfnamefont {R.~A.}\ \bibnamefont {Jones}}, \bibinfo {author}
  {\bibfnamefont {K.~W.}\ \bibnamefont {Kuper}}, \bibinfo {author}
  {\bibfnamefont {T.~B.}\ \bibnamefont {Lewis}}, \bibinfo {author}
  {\bibfnamefont {M.~T.}\ \bibnamefont {Lichtman}}, \bibinfo {author}
  {\bibfnamefont {F.}~\bibnamefont {Majdeteimouri}}, \bibinfo {author}
  {\bibfnamefont {J.~J.}\ \bibnamefont {Mason}}, \bibinfo {author}
  {\bibfnamefont {J.~K.}\ \bibnamefont {McMaster}}, \bibinfo {author}
  {\bibfnamefont {J.~A.}\ \bibnamefont {Miles}}, \bibinfo {author}
  {\bibfnamefont {P.~T.}\ \bibnamefont {Mitchell}}, \bibinfo {author}
  {\bibfnamefont {J.~D.}\ \bibnamefont {Murphree}}, \bibinfo {author}
  {\bibfnamefont {N.~A.}\ \bibnamefont {Neff-Mallon}}, \bibinfo {author}
  {\bibfnamefont {T.}~\bibnamefont {Oh}}, \bibinfo {author} {\bibfnamefont
  {V.}~\bibnamefont {Omole}}, \bibinfo {author} {\bibfnamefont {C.~P.}\
  \bibnamefont {Simon}}, \bibinfo {author} {\bibfnamefont {N.}~\bibnamefont
  {Pederson}}, \bibinfo {author} {\bibfnamefont {M.~A.}\ \bibnamefont
  {Perlin}}, \bibinfo {author} {\bibfnamefont {A.}~\bibnamefont {Reiter}},
  \bibinfo {author} {\bibfnamefont {R.}~\bibnamefont {Rines}}, \bibinfo
  {author} {\bibfnamefont {P.}~\bibnamefont {Romlow}}, \bibinfo {author}
  {\bibfnamefont {A.~M.}\ \bibnamefont {Scott}}, \bibinfo {author}
  {\bibfnamefont {D.}~\bibnamefont {Stiefvater}}, \bibinfo {author}
  {\bibfnamefont {J.~R.}\ \bibnamefont {Tanner}}, \bibinfo {author}
  {\bibfnamefont {A.~K.}\ \bibnamefont {Tucker}}, \bibinfo {author}
  {\bibfnamefont {I.~V.}\ \bibnamefont {Vinogradov}}, \bibinfo {author}
  {\bibfnamefont {M.~L.}\ \bibnamefont {Warter}}, \bibinfo {author}
  {\bibfnamefont {M.}~\bibnamefont {Yeo}}, \bibinfo {author} {\bibfnamefont
  {M.}~\bibnamefont {Saffman}},\ and\ \bibinfo {author} {\bibfnamefont {T.~W.}\
  \bibnamefont {Noel}},\ }\bibfield  {title} {\bibinfo {title} {{A universal
  neutral-atom quantum computer with individual optical addressing and
  non-destructive readout}},\ }\href {http://arxiv.org/abs/2408.08288}
  {\bibfield  {journal} {\bibinfo  {journal} {arXiv Prepr.}\ }\textbf {\bibinfo
  {volume} {2408.08288}} (\bibinfo {year} {2024})}\BibitemShut {NoStop}%
\end{thebibliography}%

\end{document}